\documentclass[aps,prb,twocolumn,floats]{revtex4}
\usepackage{color,ulem,verbatim,float,wrapfig}
\usepackage{amssymb,amsbsy,amsmath,mathrsfs}
\usepackage{appendix}

\usepackage{graphicx}
\usepackage{times}
\usepackage{color}
\usepackage[colorlinks = true,
linkcolor = blue,
urlcolor = blue,
citecolor = blue,
anchorcolor = blue]{hyperref}
\usepackage{subfigure}
\usepackage{braket}
\usepackage{bbold}
\usepackage{commath}
\usepackage{setspace}
\usepackage{bm} 
\newcommand\bea{\begin{eqnarray}}
\newcommand\eea{\end{eqnarray}}
\newcommand\beq{\begin{equation}}
\newcommand\eeq{\end{equation}}
\newcommand\bib{\bibitem}

\newcommand{\non}{\nonumber}

\newcommand{\ga}{\gamma}

\newcommand{\si}{\sigma}

\newcommand{\om}{\omega}

\newcommand{\ua}{\uparrow}
\newcommand{\da}{\downarrow}
\newcommand{\pa}{\partial}
\newcommand{\la}{\langle}
\newcommand{\ra}{\rangle}

\def\blue{\textcolor{blue}}

\newcommand\numberthis{\addtocounter{equation}{1}\tag{\theequation}}

\usepackage{soul}

\begin{document}

\title{Weak universality, quantum many-body scars and anomalous 
infinite-temperature autocorrelations in a one-dimensional spin model with duality}
\author{Adithi Udupa$^{1} \footnote{These authors contributed equally to this work}$, Samudra Sur$^{1*}$, Sourav Nandy$^2$, Arnab Sen$^3$ 
and Diptiman Sen$^{1,4}$}
\affiliation{$^1$Center for High Energy Physics, Indian Institute of Science, 
Bengaluru 560012, India \\
$^2$Jo\v zef Stefan Institute, SI-1000 Ljubljana, Slovenia\\
$^3$School of Physical Sciences, Indian Association for the Cultivation of 
Science, Jadavpur, Kolkata 700032, India \\ 
$^4$Department of Physics, Indian Institute of Science, Bengaluru 560012, India}

\begin{abstract}
We study a one-dimensional spin-$1/2$ model with three-spin interactions and
a transverse magnetic field $h$. The model is known to have a $Z_2 \times Z_2$
symmetry, and a duality between $h$ and $1/h$. The self-dual point at $h=1$ 
is a quantum critical point with a continuous phase transition. We compute the
critical exponents $z$, $\beta$, $\ga$ and $\nu$, and the central charge $c$
numerically using exact diagonalization (ED) for systems with periodic 
boundary conditions. We find that both $z$ and $c$ are equal to $1$,
implying that the critical point is governed by a conformal field theory
with a marginal operator. The values obtained for $\beta/\nu$, $\ga/\nu$,
and $\nu$ from ED suggest that the model exhibits Ashkin-Teller criticality with an effective
coupling that is intermediate between the four-state Potts model and two decoupled
transverse field Ising models. A more careful analysis on much larger systems but with open boundaries using density-matrix renormalization group (DMRG) calculations, however, reveals important additive and multiplicative logarithmic corrections at and near criticality, and we present evidence that the self-dual point may be in the same universality class as the four-state Potts model. An energy level spacing analysis shows that the model is not integrable.
For a system with an even number of sites and periodic boundary conditions,
there are exact mid-spectrum zero-energy eigenstates whose number grows
exponentially with the system size. 
A subset of these eigenstates have wave functions which are independent of 
the value of $h$ and have unusual entanglement structure; hence these 
can be considered to be quantum many-body scars. The number of such quantum scars
scales at least linearly with system size. Finally, we study the 
infinite-temperature autocorrelation functions at sites close to one end of
an open system. We find that some of the autocorrelators relax anomalously 
in time, with pronounced oscillations and very small decay rates if $h \gg 1$
or $h \ll 1$. If $h$ is close to the critical point, the autocorrelators decay 
quickly to zero except for an autocorrelator at the end site.
\end{abstract}

\maketitle

\section{Introduction}
\label{intro}

The well-known transverse field Ising model (TFIM) in one dimension has been 
studied extensively over many years~\cite{pfeuty,stinch,sachdev}. The 
Hamiltonian of the model consists of two-spin interactions (with strength set
equal to 1) and a transverse magnetic field with strength $h$, 
\begin{equation} H_2 ~=~ - ~\sum_{j=1}^{L} ~[\sigma^{z}_{j} \sigma^{z}_{j+1} ~+~
h~ \sigma^{x}_{j}], \label{eq:hamil2} \end{equation}
where $\sigma^a_j$ denote the Pauli matrices at site $j$ corresponding to a 
spin-1/2 degree of freedom, and we are considering a system with $L$ sites and 
periodic boundary conditions (PBC). The model has a $Z_2$ symmetry since an 
operator $D = \prod_{j=1}^L \si_j^x$ commutes with the Hamiltonian.
The model is known to have a quantum phase transition at a critical point given 
by $h=1$. It has a ordered phase for $h<1$ with a finite magnetization (the
$Z_2$ symmetry is spontaneously broken in this phase), and a 
disordered phase for $h>1$ with zero magnetization. It also exhibits duality 
\cite{kramers, robert} and the self-dual point $h=1$ is the quantum critical 
point. The critical point is known to be described by a conformal field theory 
with $c=1/2$ and certain critical exponents which are known 
analytically~\cite{cftbook}.

Generalizations of the TFIM with $p$-spin interactions with duality have been
proposed~\cite{turban} and studied using mean-field theory~\cite{maritan},
finite-size scaling~\cite{penson,kolb,alcaraz}, 
and series expansions~\cite{igloi}, with the TFIM
corresponding to the case $p=2$. It is of 
particular interest to take a close look at what happens in the next 
simplest case $p=3$ where the order of phase transition in literature has been 
debated. We study this case numerically using exact diagonalization (ED)
and look at the quantum criticality in this system at the self-dual point 
which is again given by $h=1$. 
Another motivation for studying the case of $p=3$ is that a Hamiltonian of this
form may be engineered using optical lattices either with two atomic
species~\cite{pachos} or with polar 
molecules driven by microwave fields~\cite{zoller}. 
We note that for the model with $p=4$, it is not clear whether the transition
at the self-dual point is first-order or continuous, while models with 
$p \ge 5$ are believed to have a first-order transition at the self-dual point~\cite{penson,maritan,igloi}.


The three-spin ($p=3$) model is a candidate for interesting high-energy behavior as well. 
For an even number of spins and periodic boundary conditions (PBC), this model satisfies 
an index theorem~\cite{michael2} that results in the presence of an exponentially large number 
(in system size) of exact mid-spectrum zero energy eigenstates. Since these states 
are degenerate in energy, any linear combination of these is also an eigenstate of the system. 
Recent works~\cite{ju,ju2,deb,kalre,sapta} have shown that this freedom allows for the possibility of 
creating mid-spectrum eigenstates which violate the eigenvalue thermalization hypothesis (ETH) 
by possessing very low entanglement entropy compared to the expected thermal entropy. 
These eigenstates can be classified as quantum many-body
scars~\cite{bernien,sanjay,turner,turner2,michael,onsagerscars,gxsu,sanada}. It would be
interesting to see if the three-spin model hosts such scar states in the 
middle of the energy spectrum. 
Finally, we would like to examine if infinite-temperature autocorrelation functions 
in open chains show anomalous behaviors as a function of time for this model. 
A motivation to do so is provided by the observation of infinite (long) coherence times for 
boundary spins for the TFIM without (with) integrability-breaking perturbations due to the 
presence of a strong zero mode (an almost strong zero mode) that commutes (almost commutes) with the 
Hamiltonian~\cite{fendley2012,fendley2016,kemp2017,yates2020,yates2020prb}.
While the TFIM can be mapped to free fermions by the standard Jordan-Wigner transformations, 
the perturbed TFIM has additional four-fermion interactions. It is not known if the three-spin model 
has analogous (almost) strong zero modes. A study of the autocorrelators
near the ends of a long system may possibly shed light on this. 

The plan of this paper is as follows. In Sec.~\ref{Ham} we present
the Hamiltonian of the model with three-spin interactions and its symmetries.
We find that the model has a $Z_2 \times Z_2$ symmetry which leads to some
degeneracies in the energy spectrum of a system with PBC. In Sec.~\ref{duality} 
we discuss the duality of the model. While the duality is easy to show
for an infinite-sized system, we discover that the existence of a duality
is a subtle issue for finite-sized systems with PBC.
In Sec.~\ref{criticality} we make a detailed study 
of the criticality properties of the model at the self-dual point given by $h=1$,
using ED for systems with PBC.
Finite-size scaling is used to first confirm that there is a critical point at
$h=1$ and then to compute the dynamical critical exponent $z$, the order parameter
exponent $\beta$, the magnetic susceptibility exponent $\ga$, and the correlation
length exponent $\nu$. We find that $z=1$ suggesting that the low-energy sector 
of the model at $h=1$ has conformal invariance. We then determine the central
charge $c$ and find that it is close to 1. Next, we
observe that although the values of $\beta$, $\ga$ and $\nu$ for the two-spin
and three-spin models are different from each other, the ratios $\beta/\nu$ and
$\ga/\nu$ are the same in the two models. This suggests that there is a weak
universality~\cite{suzuki} and the three-spin model lies on the Ashkin-Teller (AT) 
line, just like two copies of the TFIM and the four-state Potts model. Using the
numerically computed value of $\nu$ for the three-spin model, we estimate
the location of this model on the AT line of critical points.
We then perform a more careful analysis of the nature of the critical point using the density-matrix renormalization group (DMRG) method
that allows us to access much larger system sizes, but with open boundary conditions, compared to the ED method. We find evidence for important additive and
multiplicative logarithmic corrections in the critical regime which match those expected at the critical point of the four-state Potts model.
Incorporating these logarithmic corrections as well as comparing with the corresponding quantities for the quantum AT model, the data for larger chains
suggests that the self-dual point for the three-spin model may be in the same universality as the four-state Potts model.

In Sec.~\ref{scars}, we study the energy level spacing statistics to determine
if the three-spin model is integrable. We find that the level spacing statistics
has the form of the Gaussian orthogonal ensemble, and hence the model is
non-integrable. Next, we find that the model has an exponentially large number
of mid-spectrum zero-energy eigenstates. Further, we find that the zero-energy eigenstates 
are of two 
types which we call Type-I and Type-II. The Type-I states are simultaneous
zero-energy eigenstates of the two parts of the Hamiltonian (the three-spin
interaction and the transverse field) and consequently stay unchanged as a
function of $h$, thus violating the ETH. 
Hence they qualify as quantum many-body scars. We give exact expressions for a
subset of these Type-I states in terms of {\it emergent} singlets and triplets
which shows that their number increases at least linearly with system size.
In Sec.~\ref{correlator},
we study the infinite-temperature autocorrelation function at sites close to one end of a
large system and in the bulk with open boundary conditions; the purpose of this study is to 
understand if there are any states which can be interpreted as the end modes 
of a semi-infinite system. We find that
far from the critical point, at either $h \ll 1$ or $h \gg 1$, some of the
autocorrelators show an anomalous behavior in that they oscillate and also
decay very slowly with time. We provide a qualitative 
understanding of the oscillatory behavior using perturbation theory. 
For values of $h$ close to the critical point, the infinite-temperature autocorrelators decay 
quickly to zero except for a particular autocorrelator at the end site. In
Sec.~\ref{discussion} we summarize our main results and point out some 
directions for future research.

In brief, we have studied several aspects of a spin-1/2
model with three-spin Ising interactions and placed in a transverse field. The features studied include the duality and other symmetries of the model and its energy 
spectrum, the continuous phase transition and its critical exponents at the self-dual point $h=1$, a weak
universality of the critical exponents indicating that the model lies on the Ashkin-Teller line, in fact, possibly exhibiting four-state Potts universality, 
an analysis of the of energy level spacing indicating 
the non-integrability of the model, the presence of an exponentially large number of states with exactly zero 
energy, the existence of a subset of the zero energy 
states which are many-body scar states (along with an exact analytical expression for some of these scar states), and infinite-temperature autocorrelation functions near the ends of a finite-sized system which show anomalous relaxation with time.

We would like to mention here that several other one-dimensional models with 
multispin interactions have been studied over the years, and they show a
wide variety of unusual features~\cite{bonfim,fendley2019,alcaraz2}. Our 
work makes a contribution to this interesting area of research.

\section{The model and its symmetries}
\label{Ham}

The Hamiltonian of the three-spin model is given by~\cite{turban,penson,maritan,igloi,kolb,alcaraz}
\begin{equation}
H_3 ~=~ - ~\sum_{j=1}^{L} ~[\sigma^{z}_{j} \sigma^{z}_{j+1} \sigma^{z}_{j+2} 
~+~ h ~\sigma^{x}_{j}], \label{eq:hamil} \end{equation} 
where $\si_j^a$ (where $a=x,y,z$) denotes the Pauli matrices at site $j$, 
and we assume PBC so that $\si^a_{L+1}=\si^a_1$ and $\si^a_{L+2}=\si^a_2$.

\begin{figure}[H]
\centering
\includegraphics[width=0.4\textwidth]{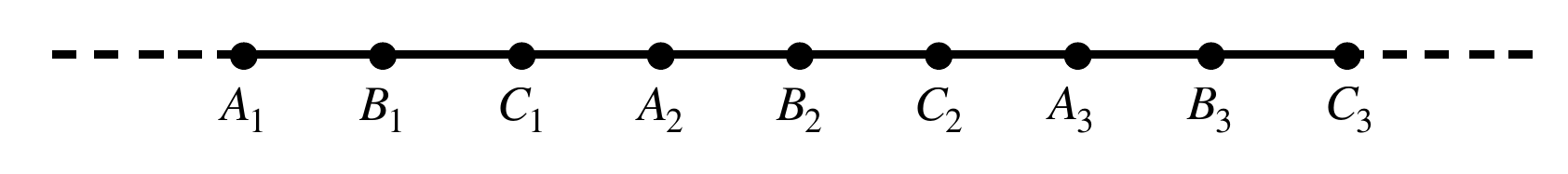}
\caption{The lattice for the Hamiltonian $H_3$ with three sublattices A, B and C are shown here. The symmetry operators are defined with respect to these sublattices in Eq.~\eqref{D_operators}.} 
\label{fig:lattice}
\end{figure}

There are three operators $D_1, D_2$ and $D_3$ which commute with the Hamiltonian $H_3$ 
in Eq.~\eqref{eq:hamil}. If the system size $L$ is a multiple of 3, we can divide the lattice into three sublattices $A$, $B$ and $C$ as shown in 
Fig.~\ref{fig:lattice}. The three operators for this system are then defined
\begin{align*}
D_1 &= \Pi_{j=1}^{L/3} \sigma_x^{A_j}\sigma_x^{B_j}, \\
D_2 &= \Pi_{j=1}^{L/3} \sigma_x^{B_j}\sigma_x^{C_j}, \\
D_3 &= \Pi_{j=1}^{L/3} \sigma_x^{C_j}\sigma_x^{A_j}. 
\numberthis
\label{D_operators}
\end{align*}
These satisfy the constraint $D_1 D_2 D_3= I$. Thus we have four decoupled sectors corresponding to the different allowed values of these operators; $(D_1, D_2, D_3) 
= (1,1,1), (1,-1,-1), (-1,1,-1)$ and $(-1,-1,1)$. Thus this model has a $Z_2 \times Z_2$ symmetry. All the four sectors have equal number of states. We also notice 
that the operator $C= \Pi_{j=1}^{L} \sigma_y$ anticommutes with the Hamiltonian. 
Hence for every state $\ket{\psi}$ with energy $E$, there is a state $C 
\ket{\psi}$ with energy $-E$ due to which the spectrum of this model has a 
$E \rightarrow -E$ symmetry. 

With PBC the system also has translation symmetry. If the translation operator
is given by $U$, then we can see from Eq.~\eqref{D_operators}, that 
\begin{align*}
U D_1 U^{-1}&= D_2, \\
U D_2 U^{-1}&= D_3, \\
U D_3 U^{-1}&= D_1.
\numberthis
\end{align*}
We can further see that a combination of these three operators $D'= D_1 + \omega D_2 + \omega^2 D_3$ where $\omega$ is the cube root of unity, transforms into $e^{-i2\pi /3} (D_1 + \omega D_2 + \omega^2 D_3)$ upon translation by one
site. This is because $U(D_1 + \omega D_2 + \omega^2 D_3) U^{-1} = \omega^2(D_1 + \omega D_2 + \omega^2 D_3)$. This means that for a state $\ket{\psi_k}$ with 
momentum $k$, that is, $U\ket{\psi_k}= e^{ik} \ket{\psi_k}$, we have a state $(D_1 + \omega D_2 + \omega^2 D_3)\ket{\psi_k}= e^{-i2\pi /3} e^{ik}\ket{\psi_k}= 
\ket{\psi_{k- 2\pi /3}}$ with momentum $k- 2\pi /3$. Similarly, we have a state $(D_1 + \omega^{-1} D_2 + \omega^{-2} D_3)\ket{\psi_k}$ for which the 
momentum is $k+ 2\pi/ 3$. Since the $D$ operators commute with the Hamiltonian, the states $\ket{\psi_k}$, $\ket{\psi_{k- 2\pi/ 3}}$ and $\ket{\psi_{k+ 2\pi /3}}$ 
are degenerate. However in the sector $(D_1, D_2, D_3) = (1,1,1)$, the operators $D_1 + \omega D_2 + \omega^2 D_3$ and $D_1 + \omega^2 D_2 + \omega D_3$ give zero
when they act on a state $\ket{\psi_k}$. Therefore the states belonging to this sector do not have a degenerate partner. Thus in the entire spectrum, 
three-fourths of the states have an exact three-fold degeneracy whereas the other one-fourth belonging to the sector $(1,1,1)$ has no degeneracy. We also have 
a parity symmetry in this system. For an even system size, we can define parity as a mirror reflection about the middle bond. The parity operator then takes the 
operator $D_1 \rightarrow D_2$ and $D_2 \rightarrow D_1$ and keeps $D_3$
unchanged. Thus, for a system with open boundary conditions which breaks translation 
symmetry, we can still have degeneracies coming from parity symmetry. These come from the states in sectors $(D_1, D_2, D_3) = (1,-1,-1)$ and $(-1,1,-1)$ as they 
go to a different sector under parity. 

\section{Duality of the model}
\label{duality}

Just like the TFIM, the three-spin model also exhibits duality on an {\it infinitely
large} system. We show this by starting from the original lattice with sites labeled by an integer $j$ which goes from $- \infty$ to $+ \infty$. 
Then the sites of dual lattice also lie at $j$. (This is
in contrast to the TFIM where the sites of the dual lattice lie at
$j+1/2$). The transformation of the Pauli matrices going from the original lattice $\sigma_j^a$ to the dual lattice $\tilde{\sigma}_j^a$ is given by
\begin{align*}
\tilde{\sigma}^x_{j+1} &= \sigma^z_{j}\sigma^z_{j+1}\sigma^z_{j+2}, \\
\tilde{\sigma}^z_{j-1} \tilde{\sigma}^z_{j}\tilde{\sigma}^z_{j+1} &= \sigma^x_j.
\numberthis
\label{transf}
\end{align*}
The Hamiltonian on the dual lattice then takes the form 
\begin{equation} \tilde{H}_3= -\sum_{j=-\infty}^{\infty} ~[ \tilde{\sigma}^{x}_{j+1} ~+~ h~ \tilde{\sigma}^{z}_{j-1} \tilde{\sigma}^{z}_{j} \tilde{\sigma}^{z}_{j+1}]. \end{equation}
Thus going from $H_3$ to $\tilde{H}_3$, the transverse field $h$ gets mapped to $1/h$. The self-dual point lies at $h= 1/h$. Hence, if $H_3$
(or $\tilde{H}_3$) has a phase transition it must occur at $|h|=1$.

We will now examine if duality also holds for a {\it finite} system with 
PBC as described in Eq.~\eqref{eq:hamil}. 
Clearly, we would like both the original and dual lattices to have the 
same number of sites, $L$, and the number of states should be $2^L$ in 
both cases. The latter can only happen if the Pauli operators are
independent operators on different sites on both the lattices. The
first equation in Eq.~\eqref{transf} and the fact that $(\si_j^z)^2 = 1$ 
for all $j$ imply that 
\bea \tilde{\sigma}^x_1 ~\tilde{\sigma}^x_2 ~\tilde{\sigma}^x_4 ~
\tilde{\sigma}^x_5 ~\cdots~ \tilde{\sigma}^x_{L-2} ~\tilde{\sigma}^x_{L-1} 
&=& I, \non \\
{\rm and} ~~~~\tilde{\sigma}^x_2 ~\tilde{\sigma}^x_3 ~\tilde{\sigma}^x_5 ~
\tilde{\sigma}^x_6 ~\cdots~ \tilde{\sigma}^x_{L-1} ~\tilde{\sigma}^x_L &=& I \eea
if $L$ is a multiple of 3. Hence there are two constraints on the $\tilde{\sigma}^x_j$ operators, implying that the eigenvalues of
the operators cannot take all possible values independently of each other.
To put it differently, the two constraints mean that the number of states
in the dual system is $2^{L-2}$ rather than $2^L$. We reach a 
similar conclusion for the original system by using the second equation 
in Eq.~\eqref{transf}. We therefore conclude that duality does not hold
for a finite system with PBC if $L$ is a
multiple of 3. It turns out that duality does hold if $L$ is {\it not}
a multiple of 3 as the Pauli operators do not satisfy any constraints
on either the original lattice or the dual lattice in that
case. (Note, however, that the operators $D_j$ defined in Sec.~\ref{Ham}
do not exist if $L$ is not a multiple of 3). Next, duality implies
that there must be a unitary operator $U_D$ which relates the states of
the original and dual lattices. 
Let us write the Hamiltonian in Eq.~\eqref{eq:hamil} in the form 
\bea H_3 &=& - ~Z~ - ~h~ X, \non \\
{\rm where}~~~ Z &=& \sum_{j=1}^L ~\si_j^z \si_{j+1}^z \si_{j+2}^z, \non \\
X &=& \sum_{j=1}^L ~\si_j^x, \label{hzx} \eea
and similarly
\beq \tilde{H}_3 ~=~ - ~\tilde{X} ~-~ h ~\tilde{Z}. \label{hdzx} \eeq
Then there must be a unitary operator $U_D$ such that $U_D X U_D^{-1} = \tilde{Z}$ 
and $U_D Z U_D^{-1} = \tilde{X}$. 
This means that at the self-dual point $h=1$, if $|\psi_n \ra$ is an
eigenstate of $H_3$ with eigenvalue $E_n$, and $|\tilde{\psi}_n \ra
= U_D | \psi \ra$
is an eigenstate of $\tilde{H}_3$ with the same eigenvalue, we must have 
\beq \la \psi_n | X | \psi_n \ra ~=~ \la \tilde{\psi}_n | \tilde{Z} |
\tilde{\psi}_n \ra ~=~ \la \psi_n | Z | \psi_n \ra, \label{virial} \eeq
where the equality $\la \tilde{\psi}_n | \tilde{Z} | \tilde{\psi}_n \ra = 
\la \psi_n | Z | \psi_n \ra$ is a consequence of self-duality. 
Since $\la \psi_n | (- X - Z) | \psi_n \ra = E_n$, Eq.~\eqref{virial} implies
that 
\beq \la \psi_n | X | \psi_n \ra ~=~ - ~\frac{E_n}{2}. \label{virial2} \eeq 
at $h=1$. A test of this relation will be discussed in Appendix~\ref{appendixC}.

Before ending this section, we note that it is not useful 
to perform a Jordan-Wigner transformation from spin-1/2's to spinless
fermions for this model because there are three-spin terms in the Hamiltonian. 
The Jordan-Wigner transformation maps $\si_j^x$ to the occupation
number $c_j^\dag c_j$ of fermions at site $j$, and $\si_j^z$ to $c_j +
c_j^\dag$ times a string of $\si_n^x$ operators running from 
$n=- \infty$ to $j-1$ (for an infinitely large system).
The presence of the three-spin term $\si_j^z \si_{j+1}^z \si_{j+2}^z$ 
in the Hamiltonian implies there will be an infinitely long string of 
$\si_n^x$ operators left over which does not cancel with anything. Thus 
this model cannot be solved by fermionizing since the fermionic 
Hamiltonian will have highly non-local terms. We will henceforth analyze 
the model numerically. In the next section, we will carry out ED calculations to
confirm the location of the critical point of the quantum phase transition 
and to extract the critical exponents.

\section{Quantum criticality of the model}
\label{criticality}

We will now study the three-spin model numerically to understand the 
nature of the phase transition at $h=1$ and the critical properties. We will 
use ED to obtain the 
ground state and low-lying excitations and then compute various thermodynamic
quantities like the magnetization and magnetic susceptibility to study the
criticality. 

\begin{figure}[H]
\centering
\includegraphics[width=0.45\textwidth]{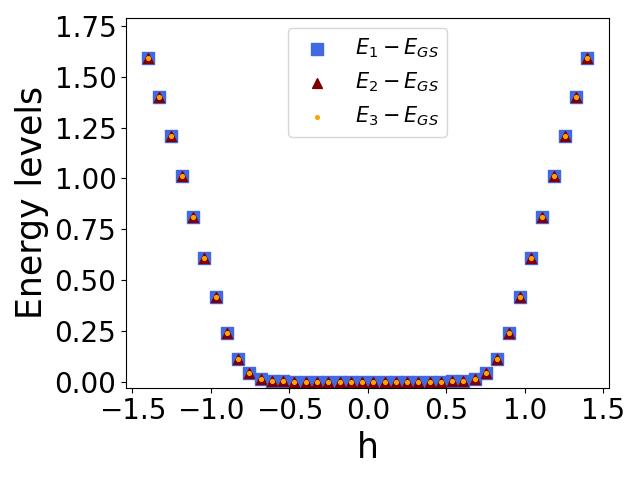}
\caption{First three energy levels as measured from the ground state energy plotted as a function of the transverse field $h$ for $L=15$. We see some
degeneracies which arise from the $Z_2 \times Z_2$ symmetry.}
\label{fig:energy_levels} \end{figure}

\subsection{Energy levels}

We use ED to compute the first few energy levels for the Hamiltonian in 
Eq.~\eqref{eq:hamil}. The first three excited energy levels with respect to the 
ground state energy are plotted in Fig.\ref{fig:energy_levels}. We first notice 
that the phase transition happens close to $|h| = 1$. In the region $ |h|>1$, the 
system is gapped with a finite difference between the ground state and the first 
excited energy. The first three excited states are exactly degenerate due to the 
symmetries $D_1, D_2, D_3$ of the model (see Sec. \ref{Ham}) with eigenvalues 
$(D_1, D_2, D_3)= (1, -1, -1), (-1, 1, -1)$ and $(-1,-1,1)$. The ground state
is unique and belongs to the sector $(D_1, D_2, D_3)= (1, 1, 1)$. In the region 
$|h| <1$, the ground state becomes degenerate with the three-fold degenerate states 
as the system size approaches infinity. For finite-sized systems, there
is a small gap in the region $|h|<1$. The gap varies with $h$ and falls off 
exponentially with the system size; for $h=0.4$ and $L=15$, the gap is of the order 
of $10^{-4}$. 

\subsection{Finite-size scaling}

To understand the nature of the phase transition in this model in comparison to the 
TFIM which has two-spin interactions, we look at the behaviors of different 
quantities close to the critical point. Close to the critical point, any 
singular quantity, $\Theta$, will have an asymptotic behavior of the 
form~\cite{pang} 
\beq \Theta \sim |h - h_c|^{-\theta}, \label{eq:expo} \eeq
where $\theta$ is the critical exponent of the quantity $\Theta$.
In addition, continuous phase transitions have a diverging correlation length
scale $\xi$ which diverges close to the critical point as $\xi 
\sim |h- h_c|^{-\nu}$,
where $\nu$ is the critical exponent corresponding to the correlation length. 
This implies that $\Theta \sim \xi^{\theta/\nu}$.
At the critical point, the correlation length diverges. However for finite system sizes, we are limited by the system size $L$. Hence, when the correlation length exceeds the system size, the quantity will vary with $L$ depending on the ratio $L/\xi$, and the above relation gets modified to 
\beq \Theta \sim \xi^{\theta/\nu} \Theta_0(L/\xi),
\label{eq:omega} \eeq
where $\Theta_0(L/\xi)$ is a scaling function with 
\begin{center}
\[ \Theta_0(L/\xi)=
\begin{cases}
\text{constant} & {\rm for} ~~~L \gg \xi \\
(L/\xi)^{\theta/\nu} & {\rm for} ~~~L \ll \xi.
\end{cases}
\]
\end{center}


Thus at the critical point when $\xi \gg L$, we find that $\Theta$ scales as \cite{pang}
\beq \Theta |_{h_c} \sim L^{\theta/\nu}. \label{eq:scale} \eeq
By evaluating $\Theta$ for different system sizes we can calculate the critical exponent $\theta/\nu$ once we know the exact location of the critical point 
$h_c$.

\subsection{Numerical determination of critical point}
\label{sec:fidresults}

The ground state fidelity is the one of the preliminary ways to detect a quantum phase transition. The fidelity is defined as
$\mathcal{F} (h,\delta h) = |\langle \psi_0 (h - \delta h /2) \mid \psi_0 (h+ \delta h /2) \rangle|$, where $\psi_0 (h \pm \delta h/2)$ is the ground state of the Hamiltonian with parameter $h \pm \delta h/2$, and $\delta h$ is a small but fixed number. The fidelity is expected to show a pronounced deviation from unity in the neighborhood of a
phase transition.

\begin{figure}
\centering
\includegraphics[width=0.45\textwidth]{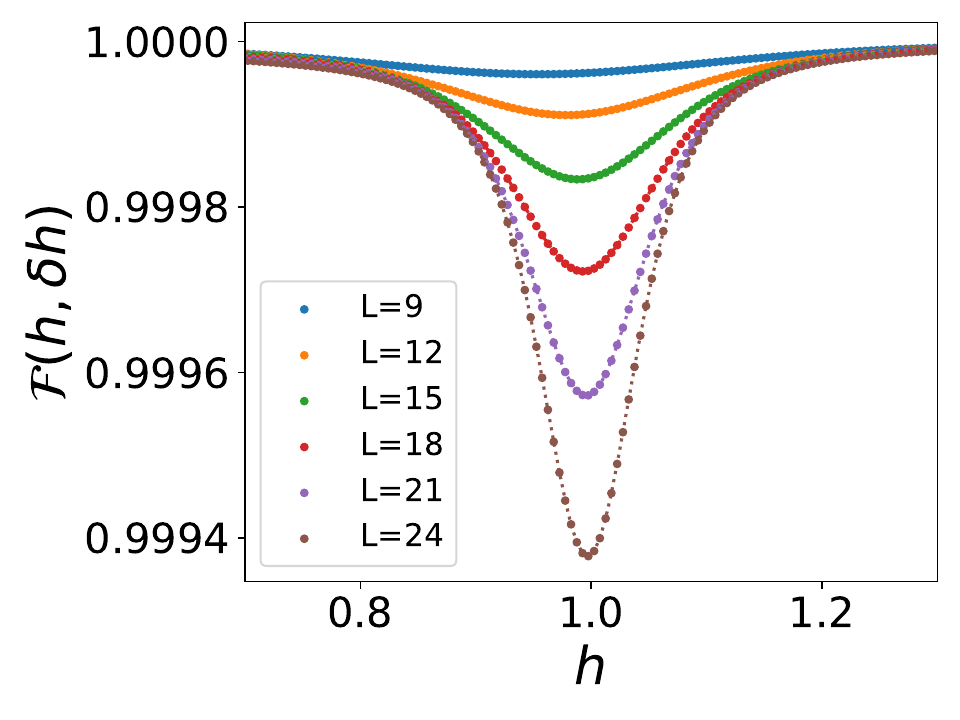}
\caption{Fidelity as a function of the transverse field $h$ (for a fixed
$\delta h=0.005$) is plotted for different system sizes $L=9,12,15,18,21$ and 
$24$. There is a dip in the fidelity close to the expected critical point $h=1$.} 
\label{fig:fidelity} \end{figure}

In Fig.~\ref{fig:fidelity} we show the variation of the fidelity $\mathcal{F}(h, \delta h)$ 
as a function of the transverse field $h$ for different system sizes $L= 9, 12, 15, 
18, 21, 24$ for a fixed $\delta h=0.005$. We see a dip close to $h = 1$ for all system sizes confirming that a 
phase transition occurs at this point. As the system size increases, the magnitude 
of the dip increases and the location of the dip approaches the predicted value 
$h=1$. For the largest system size here $L=24$, we find that the minimum occurs
at $h_c= 0.9960$. We also note that the location of the minimum, $h_c(L)$,
obtained from the fidelity 
scales as $h_c(L)=1+aL^{-2/\nu}$, while the value of the
fidelity susceptibility
$\chi_F(h_c=1)=-\partial^2 \mathcal{F}(h_c,\delta h)/\partial^2 (\delta h)|_{\delta h \rightarrow 0}$ at $h_c=1$ scales as $bL^{2/\nu}$,
where $a,b$ are constants and $\nu$ is 
the correlation length exponent yielding
$\nu \approx 0.71$ (see Sec.~\ref{nu} for further discussion). 


\subsection{Dynamical Critical Exponent $z$}
\label{z}

The smallest energy gap in the system at finite sizes (Fig.~\ref{fig:energy_levels}) can be used to estimate the dynamical critical exponent $z$. As we approach the critical point, the energy difference between
the first excited state and the ground state, $\Delta$, behaves as 
\beq \Delta ~\sim~ |h- h_c|^{z\nu}. \eeq
Given the exponent $z\nu$, Eqs.~\eqref{eq:expo} and \eqref{eq:scale} imply that
\beq \Delta|_{h_c} ~\sim~ L^{-z}. \label{eq:z}\eeq
We evaluate $\Delta$ by performing ED for various system sizes $L= 12, 15, 21, 24$ and $27$ in the neighborhood of the critical point.
Fig.~\ref{fig:est_c} (a) shows the variation of $\Delta$ with $h$ for different system sizes. At $h=h_c$ we plot a log-log graph of $\Delta|_{h_c}$ versus $L$ (inset of Fig.~\ref{fig:est_c} (a)), and fit it linearly to obtain the slope. We find that $z= 1.0267 \pm 0.0014$ indicating that $z=1$ at criticality.

\subsection{Calculation of central charge $c$}
\label{c}

\begin{widetext}
\begin{figure*}
\centering
\includegraphics[width=0.9\textwidth]{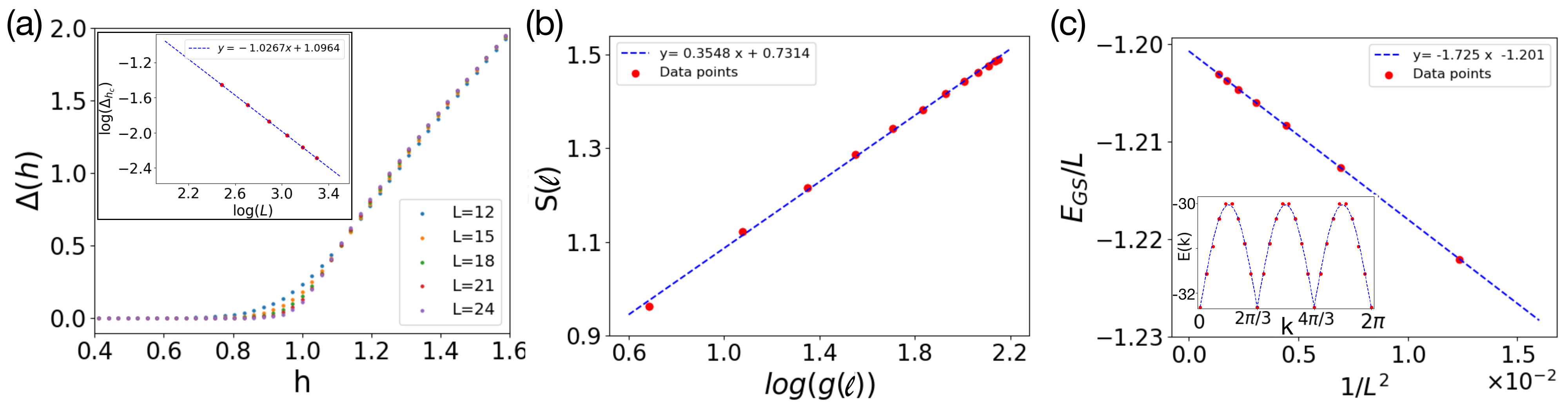}
\caption{(a) Plot of the smallest gap, $\Delta$, as a function of $h$ for
different system sizes. The inset shows a log-log plot of $\Delta|_{h_c}$ fitting which to a straight line gives the dynamical exponent $z=1.0267 \pm 0.0014$. (b) Plot of the ground state entanglement entropy versus the logarithm 
of $g(l)= (L/\pi) \sin (l\pi/L)$, where $l$ is the size of one of the subsystems, and $L=27$ at the critical coupling $h_c=1$. The slope of the graph is $c/3$ which gives $c= 1.0644$.
(c) From Eq.~\eqref{GSscaling}, the variation of the ground state energy with the 
system size $L$ at $h_c=1$ gives an estimate for $c$, namely, the slope of $E_{GS}/L$ versus $1/L^2$ has a slope equal to $-\pi v c/(6L)$. For $L= 27$, 
we find that $c=0.9585$. The inset shows the velocity estimate of the gapless excitations which is calculated by fitting the function $E(k)= a \sin (b k) + d$. For $L=27$, we find $a = 2.2893$ and $b= 1.5012$
respectively giving the velocity $v= ab = 3.4367$. } \label{fig:est_c} \end{figure*}
\end{widetext}

Since the critical exponent $z=1$ for this model, the low-lying excitations at the 
critical point have a linear dispersion making 
the system Lorentz invariant with some velocity $v$ which will be discussed
below. Thus the model can be described by a $1+1$-dimensional conformal field
theory characterized by a central charge $c$~\cite{cftbook}. 
In such a theory, the von Neumann entanglement entropy of the system can be used to 
extract the central charge $c$. If the system is divided into two subsystems A and
B, the von Neumann entanglement entropy between the two systems is given by 
\beq S_A ~=~ -\text{Tr}_A (\rho_A \text{log}\rho_A), \eeq
where $\rho_A$ is the reduced density matrix of the subsystem A obtained by 
tracing out the states in B from the density matrix of the ground state: $\rho_A = 
Tr_B \ket{\psi_{GS}}\bra{\psi_{GS}}$. For a finite system size $L$ with PBC, if 
we divide the system into two subsystems with sizes $l$ and $L-l$, the von Neumann 
entanglement entropy for the subsystem $l$ is found to be~\cite{cardy}
\beq S(l) ~=~ \frac{c}{3} ~\ln [g(l)] ~+~ c', \eeq
where $g(l)= (L/\pi) \sin (\pi l/L)$, and $c'$ is a constant. For our model,
we take $L=27$ and calculate $S(l)$ for different subsystems $l$, plot $S(l)$ 
(Fig.~\ref{fig:est_c} (b)) as a function of $\ln [g(l)]$, and fit it linearly. The central charge $c$ is three times the slope obtained from this fit which gives $c= 1.0644 \pm 0.0072$. 

We can use another method to calculate $c$.
The ground state energy of a finite-sized system is found to show the following dependence on the system size $L$~\cite{cardy},
\begin{equation} E_{GS}= \alpha L- \frac{\pi v c}{6L}, \label{GSscaling} \end{equation} 
where $\alpha$ is a non-universal constant equal to the ground state energy per
site in the thermodynamic limit \cite{chen}, $v$ is the velocity of the gapless 
excitations at the critical point which can be obtained from the dispersion, and 
$c$ is the central charge. We first calculate the velocity by plotting the 
dispersion for $L=27$ as shown in the inset of Fig.~\ref{fig:est_c} (c). As discussed earlier, the dispersion varies periodically with the momentum with a
period equal to $2\pi/3$. Fitting the inset in Fig.~\ref{fig:est_c} (c) with a function of the form $E = a \sin (b k) + d$, where
$a= 2.2893 \pm 0.0134$ and $b = 1.5012 \pm 0.0015$ 
respectively (the value of $b$ is consistent with a period of $2 \pi/3$. 
Thus the velocity in the linear region near $k=0$ is $v= a b = 3.4367 \pm 0.0236$. 
The slope of $E_{GS}/L$ versus $1/L^2$ shown in Fig.~\ref{fig:est_c} (c) gives a 
slope equal to $-\pi v c/(6L)$. Putting all this together, we get the value of 
$c$ for this model to be $c= 0.9585 \pm 0.0015$. Thus both the methods 
give an estimate of $c$ which is close to $1$. A value of $c = 1$ suggests
the possibility of a marginal operator at the critical point~\cite{verlinde} of the three-spin
model, and hence weak universality.
To investigate this further, we proceed to compute the other critical 
exponents of this system: $\beta$ related to the order parameter,
$\gamma$ to the 
magnetic susceptibility, and $\nu$ to the correlation length.

\subsection{Order parameter exponent $\beta$}
\label{beta}

We now study the order parameter in this model. Given the three-spin form of the interaction, we define a symmetric order parameter as follows. As described earlier, the lattice has three sublattices $A, B$ and $C$.
We define three quantities
\begin{align*}
m_A &= \frac{3}{L} ~\sum_{n=1}^{L/3} ~\sigma^z_{3n-2}, \\
m_B &= \frac{3}{L} ~\sum_{n=1}^{L/3} ~\sigma^z_{3n-1}, \\
m_C &= \frac{3}{L} ~\sum_{n=1}^{L/3} ~\sigma^z_{3n},
\numberthis
\end{align*}
and a combined order parameter
\begin{equation} m= \sqrt{\langle m_A^2\rangle+ \langle m_B^2\rangle + 
\langle m_C^2\rangle}. \label{order_parameter} \end{equation}

For numerical clarity, it would be worthwhile to note here that for finite-size systems, the ground state expectation values $\langle m_a \rangle$ are equal to
zero for $a=A,B,C$ even for $h<h_c$. This is due to the fact that the ground 
state is 
four-fold degenerate (in the infinite size limit), and the ground state obtained from ED is a linear combination of these four states making the expectation values exactly equal to zero. To bypass this problem we have first evaluated $\langle m_a^{2}\rangle$ and then taken the square root of the squares.
The behavior of 
$m$ for our model as a function of transverse field $h$ is shown in Fig.~\ref{fig:critical_exp} (a). It begins to drop to zero as we approach $h_c$. Close to the critical point, we have
\beq m \sim |h-h_c|^\beta. \eeq
From the finite-size scaling of magnetization, we have
\beq \mathcal{M}_z\sim L^{-\beta/\nu}, \eeq
where $\mathcal{M}_z = m|_{h_c}$. The log-log graph for $\mathcal{M}_z$ 
versus $L$ is shown in the inset of Fig.~\ref{fig:critical_exp} (a);
from this we find that $\beta/\nu= 0.1291 \pm 0.0018$. This ratio is close 
to the value of $\beta/\nu=1/8$
found for the TFIM (two-spin model) where it is analytically known 
that $\beta= 1/8= 0.125$ and $\nu=1$.

\subsection{Magnetic susceptibility exponent $\gamma$}
\label{expgamma}

We now compute the magnetic susceptibility $\chi$. For this calculation, we add a longitudinal field to the system so that the Hamiltonian becomes 
\beq H ~=~ - ~\sum_{j=1}^{L} ~[\sigma^{z}_{j} \sigma^{z}_{j+1} \sigma^{z}_{j+2} 
~+~ h~ \sigma^{x}_{j} ~+~ h_z~ \sigma^{z}_{j}], \label{eq:hamil3} \eeq
where $h_z$ is the longitudinal field in the system.

\begin{widetext}
\begin{figure*}
\centering
\includegraphics[width=\textwidth]{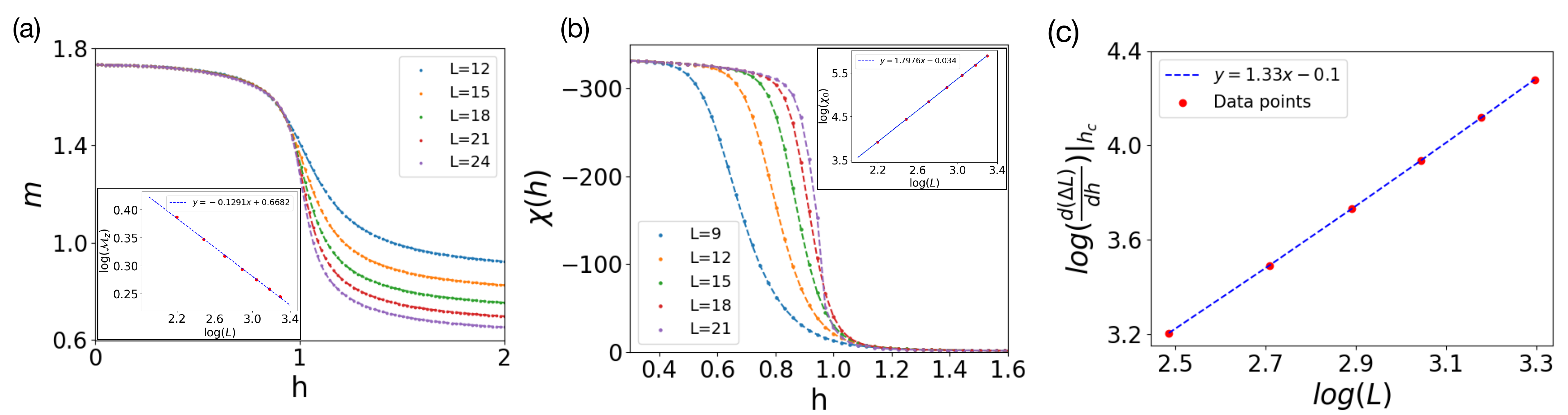}
\caption{(a) Plot of the order parameter defined in Eq.~\eqref{order_parameter} versus $h$ for different system sizes. We see that
it has a finite value for $h<1$ and falls off as $h>1$. The log-log plot of this quantity at the critical point with maximum system size $L=27$ gives a 
slope of $\beta/ \nu$ close to $0.129$. (b) Plot of the magnetic susceptibility for this model as a function of $h$. At the critical point, it scales with the 
system size with an exponent $\gamma$. From the log-log plot shown in the inset, we find that $\gamma/\nu = 1.7976$. (c) A log-log plot of 
$d(\Delta(h) L)/dh|_{h_c} $ at the critical point versus $L$. The slope gives us the inverse of the exponent of correlation length $\nu$. From this data 
fitting, we obtain the value of $\nu$ to be 0.7538.} \label{fig:critical_exp}
\end{figure*}
\end{widetext}

The magnetic susceptibility is defined as~\cite{um}
\beq \chi = \frac{\pa{\langle M_{h_c} \rangle}}{\pa{h_z}}|_{h_z \rightarrow 0},
\eeq
where $\langle M_{h_c} \rangle$ is computed as follows. 
We first define $M= \frac{1}{L} \sum_{i=1}^L \sigma^z_i$ and evaluate its expectation value in the ground state as a function of the transverse and
longitudinal fields $h_z$ and $h$. It will be non-zero due to the 
presence of the longitudinal field. At the critical point $h_c=1$ we take the 
derivative of $M_{h_c}$ with respect to $h_z$ and find its value in the limit 
$h_z \to 0$. The magnetic susceptibility as a function of the transverse 
field $h$ is shown in Fig.~\ref{fig:critical_exp} (b). 

For different system sizes at the critical point we have the quantity $\chi_0 = \chi|_{h_c}$ which, from finite-size scaling, behaves as
\beq \chi_0 \sim L^{\ga/\nu}, \label{gamma} \eeq
where $\ga$ is the exponent corresponding to susceptibility. This is estimated by plotting a log-log graph of $\chi_0$ versus $L$ as shown in the inset of 
Fig.~\ref{fig:critical_exp} (b). The ratio of the exponents $\ga/\nu$ comes
out to be $1.7976 
\pm 0.0034$ for this model, which is close to the value of $\ga/\nu=7/4$
known for the TFIM where $\ga = 7/4=1.75$ and $\nu=1$. 

\subsection{Correlation length exponent $\nu$}
\label{nu}

To evaluate the correlation length exponent, we return to Eq.~\eqref{eq:omega}. Reorganizing that equation using the relation $\tilde{\Theta}(y) = y^{\theta}\Theta_0(y^{\nu})$, we get 
\beq \Theta L^{-\theta/\nu}  \sim \tilde{\Theta}(L^{1/\nu}|h-h_c|).
\label{eq:collapse} \eeq
By choosing the thermodynamic quantity $\Theta$ to be the smallest energy gap $\Delta$ and by fixing the corresponding exponent $z$ to be 1 as obtained 
earlier, we look at the derivative of Eq.~\eqref{eq:collapse} with respect to $h$. At the critical point, we therefore have 
\beq \frac{d(\Delta L^{z})}{dh}|_{h_c} \sim L^{1/\nu}.
\label{eq:derivative} \eeq
In Fig.~\ref{fig:critical_exp} (c), we plot the logarithm of $\frac{d(\Delta L^{z})}{dh}|_{h_c}$ versus the logarithm of the system sizes 
with $L=$ 12, 15, 18, 24 and 27. From the slope of the data fitting, we find the value of $\nu= 0.7538$. From this analysis, we see that the critical point of the three-spin model is
different from the TFIM (where $\nu=1$) even though the values of $\beta/\nu$
and $\gamma/\nu$ seem to be identical.

\subsection{Comparison with transverse field Ising model,
hyperscaling, and quantum Ashkin-Teller model}

We have repeated the numerical analysis for the TFIM (two-spin Ising model)
using ED for system sizes $L= 8, 10, 12, 14, 16, 18, 20$ and $22$.
In that case our calculations give $z= 1.0026$, $c \approx 0.50$,
$\beta /\nu = 0.1337$ and
$\ga/\nu=1.7936$.
For 
the three-spin model we found above that $z= 1.02$, $\beta /\nu = 0.129$ and 
$\ga/\nu=1.798$ with the data from system sizes $L= 9, 12, 15, 18, 21, 24, 27$.
We can see that the values of the ratios of critical exponents
$\beta/\nu$ and $\gamma/\nu$ are very close to each other for the 
two models. However the correlation length critical exponent $\nu$ is $1$ for
the two-spin model (TFIM) and close to $0.75$ for the three-spin model.
Since all the
exponents and the central charge value conform with the theoretical values from 
analytical and numerical calculations for 
the two-spin Ising model~\cite{friedan, pang}, we expect that the exponents 
obtained by the same methods for the three-spin model are also reliable. The
estimated values for the two models are tabulated and compared in Table 
\ref{table1}. Furthermore, we check for the validity of the hyper-scaling 
relation for our model. The hyperscaling relation is given by \cite{pang}
\beq 2\beta +\ga ~=~ \nu(d+z), \label{hyper} \eeq
where $d$ is the space dimensionality of the system ($d=1$ in our case). 
Since $d$, $z$, $\beta /\nu$ and $\ga /\nu$ are the same for the two-spin 
and three-spin models, our model also satisfies the hyperscaling relation. 

Since the central charge $c=1$ for the three-spin model and the ratios of the
critical exponents $\beta/\nu$ and $\gamma/\nu$ are essentially identical to those of the TFIM, this strongly suggests that the critical behavior of the
three-spin model belongs to the class of $1+1$-dimensional models with
$z=1$ and $c=1$ described by the AT model~\cite{ashkin}.
The AT model constructed on a lattice has two spin $s=1/2$
freedom on each site denoted by $\sigma$ and $\tau$. These operators 
are coupled by a parameter $\lambda$. The Hamiltonian for the quantum AT 
model is given by~\cite{chepiga}
\bea H_{AT} &=& - ~\sum_{j=1}^{L} ~( \sigma_j^z \sigma_{j+1}^z + \tau_j^z \tau_{j+1}^z +
\lambda ~\sigma_j^z \sigma_{j+1}^z \tau_j^z \tau_{j+1}^z) \non \\
&& - ~h ~\sum_{j=1}^L ~( \sigma_j^x + \tau_j^x +\lambda ~\sigma_j^x \tau_j^x). \label{eq:ATdef} \eea
For any value of $\lambda$ and $h$, this model has a $Z_2 \times Z_2$ 
symmetry, where the two $Z_2$'s are given by $(\si_j^z \to - \si_j^z, ~\si_j^x \to \si_j^x)$ and
$(\tau_j^z \to - \tau_j^z, ~\tau_j^x \to \tau_j^x)$ respectively.

\begin{table}[H]
\begin{center}
\begin{tabular}{||c| c| c| c||} 
\hline
Exponent & ~Method used~ & ~Three-spin~ & ~Two-spin~ \\ [0.5ex] 
\hline
 $z$ & $\Delta$ scaling with $L$ at $h_c$ & 1.0267 (14) & 1.0026 (3)~ \\ [1ex] 
 \hline
 $\beta$ & $m$ scaling with $L$ at $h_c$ & 0.0973 (14) & 0.1337 (64)~ \\ [1ex] 
 \hline
 $\ga$ & $\chi$ scaling with $L$ at $h_c$ & 1.3550 (84) & 1.7936 (20)~ \\ [1ex] 
\hline
$ \nu$&  $\dfrac{d}{dh}(\Delta L)$ scaling with $L$  & 0.7538 (45) & 1.0335 (42) \\ [1ex] 
& at $h_c$ & & ~ \\ [1ex]
\hline
$c$ & EE Scaling at $h_c$& 1.0644 (72) & 0.5096 (13)~ \\ [1ex] 
& Energy scaling at $h_c$ & 0.9585 (15) & 0.5034 (68)~ \\ [1ex] 
\hline
\end{tabular}
\end{center}
\caption{Numerical estimates of the critical exponents and the central charge for the three-spin and
two-spin Ising models in a transverse field. Here EE stands for entanglement entropy. The error bars shown are obtained from the fitting procedures as discussed in the text.} \label{table1}
\end{table}

At the critical point $h=1$, the model in Eq.~\eqref{eq:ATdef} is known to exhibit weak universality, namely, the ratios of the 
exponents $\beta/\nu = 1/8$ and $\gamma/\nu = 7/4$ are independent of 
$\lambda$ but the values of the exponents individually depend on $\lambda$. 
One limit of $\lambda=0$ reduces the AT model to two decoupled TFIM 
thus giving $c=1$. For this case we know that $\nu=1$. In the other limit of 
$\lambda=1$, we get the four-state Potts model~\cite{wu}, with the critical 
exponent $\nu=2/3$. We thus see that our three-spin model $H_3$ also appears to
show this weak universality since $c=1$, and $\beta/\nu$ and $\gamma/\nu$ are
close to $1/8$ and $7/4$. However $\nu$ is different from the TFIM. 
Since the value of $\nu= 0.7538$ for the three-spin model, results from ED suggest that it must lie 
somewhere in between two copies of the the TFIM and the four-state 
Potts model. To find the value of $\lambda$ for which the three-spin model 
would get mapped to the AT model, we would have to study the AT model as a
function of $\lambda$. However since the number of degrees of freedom are doubled, we can go only up to system sizes $L=13$ using ED, and thus cannot 
rely on those numerical results. We note that the value of $\nu \approx 0.75$ for the three-spin model is consistent
with the values obtained earlier by finite-size scaling for system sizes up to
$L=18$~\cite{penson,kolb,alcaraz} and by series expansions~\cite{igloi}.

\subsection{DMRG Results}
\label{dmrg}
We now present the results extracted from studies using the density-matrix
renormalization group (DMRG) method \cite{White_DMRG, Ulrich_rev} for 
the three-spin model with open boundary conditions (OBC). The primary objective of performing the DMRG is to better understand the nature of the critical point 
at $h_c=1$. We implement DMRG using the ITensor (Julia) library \cite{Itensor}. The reasons for using OBC while performing DMRG studies are twofold: (i) the 
traditional DMRG algorithm (and the one used by the ITensor library) is based on optimization of open matrix product states \cite{Itensor2}, and (ii) even 
though one 
can try in a straightforward manner to implement DMRG with PBC simply by including a `long bond' directly connecting the two ends of the systems, such a naive 
approach suffers from serious drawbacks. In particular, if DMRG with OBC achieves a certain accuracy while keeping 
$\chi$ states (where $\chi$ denotes the bond dimension), then to reach the same accuracy with PBC, one needs to keep approximately $\chi^{2}$ states 
\cite{Itensor2}. This drawback of the DMRG method with PBC makes it rather costly and therefore impractical. We note that although there are certain proposals for 
efficient DMRG studies with PBC \cite{Pippan}, such proposals are yet to be well tested for critical systems to the best of our knowledge.

\begin{widetext}
\begin{figure*}
\centering
\includegraphics[width=0.95\textwidth]{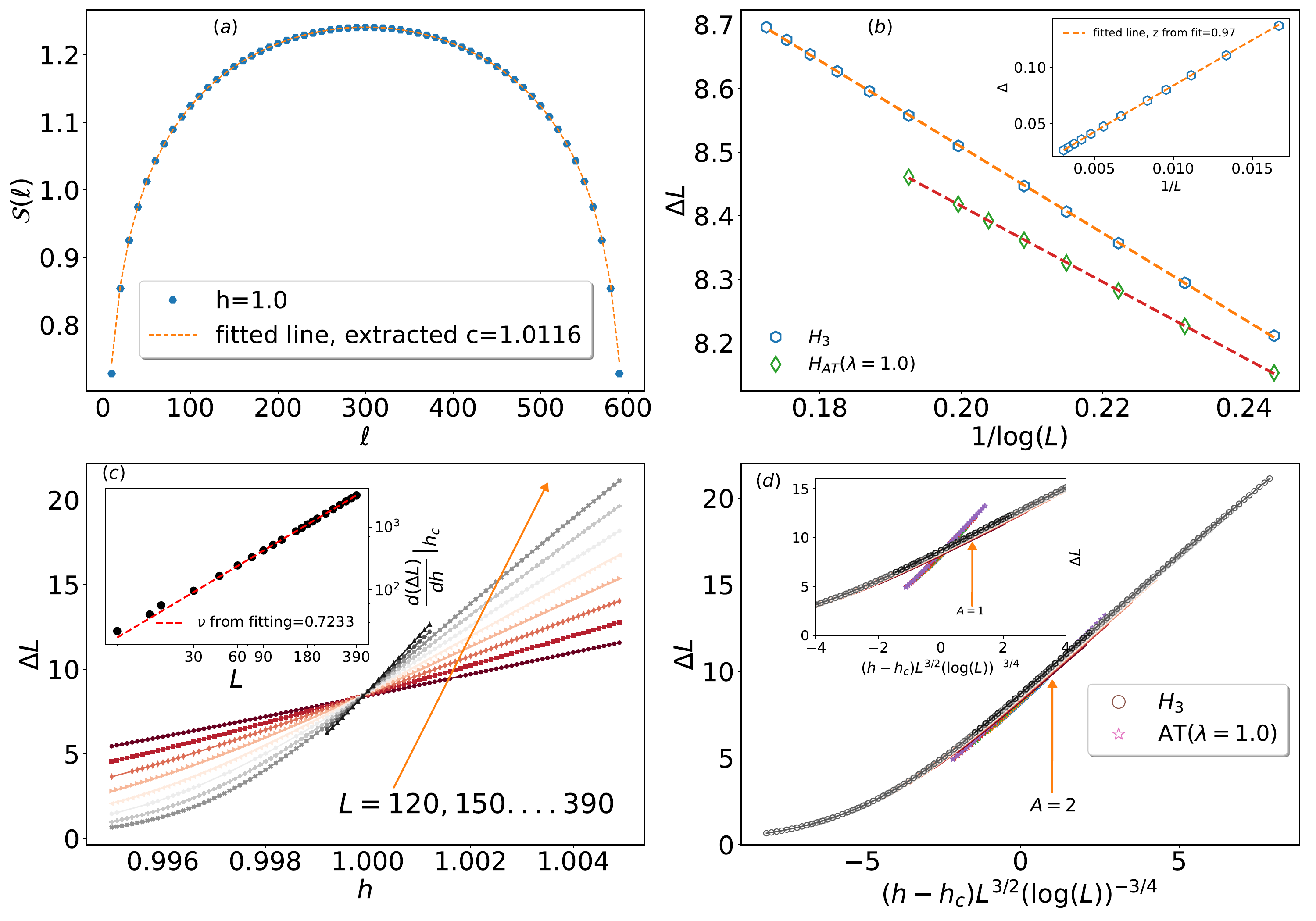}
\caption{(a) Behavior of the entanglement entropy $\mathcal{S}(\ell)$ as a function of the subsystem size $\ell$ in the ground state of the three-spin 
model at the critical point $h_c=1$ for $L=600$. (b) Behavior of $\Delta L$ as a function of system size for both the three-spin model and the AT model at 
$\lambda=1$ is consistent with the form shown in Eq.~\eqref{eq:additivelog}. The inset shows the result of fitting the corresponding data for the three-spin 
model to the form $\Delta|_{h_c} ~\sim~ L^{-z}$. (c) Plot of $\Delta L$ versus $h$ in the vicinity of $h_c=1$ for various $L$ shown for the three-spin model. 
The inset shows the determination of the exponent $\nu$ using $\frac{d(\Delta L)}{dh}|_{h_c}\sim 
L^{1/\nu}$. (d) $\Delta L$ in the neighborhood of $h_c$, both for the three-spin model and the AT model at $\lambda=1$, can be collapsed to the same 
universal curve by assuming the scaling form shown in Eq.~\eqref{eq:collapselog} and choosing the metric factor $A=1$ in the former 
case and $A \approx 2$ in the latter case. The inset shows the scaling collapse assuming Eq.~\eqref{eq:collapselog} without adjusting for the different metric 
factors in the two different models.}
\label{fig:dmrg} \end{figure*}
\end{widetext}

Using DMRG, we first compute the entanglement entropy $\mathcal{S}(\ell)$ as a function of the subsystem size $\ell$ in the ground state of the three-spin model at 
the critical point $h_c=1$. We extract the central charge $c$ using the well-known relation $\mathcal{S}(\ell) = 
(c/6)\ln ((L/\pi a)\sin(\pi \ell/L)) + c^{\prime}$ \cite{cardy} on open chains. As shown in Fig. \ref{fig:dmrg} (a), using data for a 
system size of $L=600$, we 
get $c=1.0116 \pm 0.00071$ (along with $c^{\prime}=0.3552 \pm 0.00058$) which provides strong evidence that the $h_c=1$ critical point is described by a $c=1$ conformal field theory.
Furthermore, we can reliably extract the smallest excitation gap above the ground state, $\Delta$, in the neighborhood of $h_c=1$ for open chains up to $L 
\leq 390$ which gives us a way to estimate both $z$ and $\nu$. We also calculate the smallest gap, $\Delta$, in the vicinity of the critical point 
$h_c=1$ for the AT model at various values of $\lambda$ (Eq.~\eqref{eq:ATdef}) to compare certain universal features at criticality with the three-spin model. Since the 
local Hilbert space for the AT model is four-dimensional unlike that of the three-spin model, we can only go up to open chains of size $L \leq 180$ with 
our available resources.

As in the ED procedure, we first estimate $z$ using $\Delta|_{h_c} ~\sim~ L^{-z}$ (see the discussion around Eq.~\eqref{eq:z}). For the three-spin model, 
using data for $L \in [90,330]$, we obtain $z \approx 0.97$ which is still reasonably far from the theoretically expected value of $z=1$ (see the inset of 
Fig.~\ref{fig:dmrg} (b)). In contrast, we obtain $z \approx 0.995$ for the AT model at $\lambda=0$ (the decoupled Ising limit) even after using smaller 
system sizes $L \in [75, 180]$. On the other hand, the analysis of $\Delta$ for the critical point of the AT model in the vicinity of $\lambda=1$ again yields 
$z \approx 0.97$ with the available system sizes. This suggests the presence of slowly decaying corrections in $L$ for $\Delta|_{h_c} L$. It is well-known from 
the study of the critical properties of the classical two-dimensional four-state Potts model~\cite{Cardy1980,Salas1997} that there are important 
universal additive and multiplicative logarithmic corrections to scaling in many quantities. From the classical-to-quantum correspondence, we expect the 
same in the one-dimensional quantum version. In fact, one expects a leading additive logarithmic correction of the form~\cite{Salas1997}
\beq \frac{\Delta|_{h_c} L}{|h_c|} ~=~ a^* ~+~ \frac{b}{\ln (L)} ~+~ 
\cdots, \label{eq:additivelog} \eeq
where $a^*$ is a universal number that characterizes the critical point and $\cdots$ refers to terms that decay faster than $1/\ln (L)$. We indeed see 
from Fig.~\ref{fig:dmrg} (b) that 
the smallest gap $\Delta|_{h_c}$ for both the three-spin model and the AT model at $\lambda=1$ are consistent with this form. We 
choose to only consider the leading additive logarithmic correction while fitting the data since we only have a limited range of $L$ available to us from 
DMRG. We estimate $a^* \approx 9.86 (9.61)$ from the data for the three-spin model (AT model at $\lambda=1$) from the available system sizes.

We estimate $\nu$ from the DMRG data by using the relation $\frac{d(\Delta L)}{dh}|_{h_c}\sim 
L^{1/\nu}$ (see the discussion around Eq.~\eqref{eq:derivative}) and fitting system sizes in the range of $L \in [90, 390]$ for the three-spin model, from which we 
extract $\nu=0.7233 \pm 0.0011$ as shown in the inset of Fig. \ref{fig:dmrg} (c). In the main panel of the same figure, we show the behavior of $\Delta L$ versus $h$ 
for different system sizes. We obtain a good scaling collapse (not shown) when $\Delta L^{z}$ (with $z=1$) is 
plotted as a function of $(h-h_c)L^{1/\nu}$ (see the discussion around Eq.~\eqref{eq:collapse}), with $\nu \approx 0.72$. The significant difference 
between the numerical values of $\nu$ obtained via ED ($\approx 0.75$) and 
DMRG plausibly indicates the 
strong sensitivity of the system to both finite size effects and the boundary conditions. However, we note that $\frac{d(\Delta L)}{dh}|_{h_c}\sim 
L^{1/\nu}$ follows from Eq.~\eqref{eq:collapse}. On the other hand, it is known that there are important multiplicative logarithmic corrections to the scaling 
form (Eq.~\eqref{eq:collapse}) for $\Delta$ in the vicinity of the critical point of the two-dimensional classical four-state Potts model. In fact, we 
expect that
\beq \Delta L ~=~ \mathcal{F} (A(h-h_c)L^{3/2}(\ln L)^{-3/4})
\label{eq:collapselog} \eeq 
from the behavior of the correlation length near criticality for the two-dimensional classical four-state Potts model~\cite{Salas1997}, where 
$\mathcal{F}$ denotes a universal function and $A$ denotes a {\it metric factor} that varies depending on the microscopics of the model. Assuming 
Eq.~\eqref{eq:collapselog} gives
$\frac{d(\Delta L)}{dh}|_{h_c}\sim L^{3/2}(\ln L)^{-3/4}$ which {\it mimics} a power law $L^{1/\nu}$ with $\nu \approx 0.72$ given that $L \in [90, 390]$. In fact, the 
inset of Fig. \ref{fig:dmrg} (d) shows that the $\Delta$ in the neighborhood of $h_c=1$ for different $L$ is consistent with this scaling ansatz 
(Eq.~\eqref{eq:collapselog}) both for the three-spin model and for the AT model with $\lambda=1$ (where $L \in [75, 180]$). Note that the inset of Fig. 
\ref{fig:dmrg} (d) requires no fitting parameter. In fact, choosing a metric factor of $A=1$ for the three-spin model and $A \approx 2$ (by visual 
inspection) for the AT model at $\lambda=1$ makes the data for the two different microscopic models fall on the same scale-collapsed curve 
(Fig.~\ref{fig:dmrg} (d), main panel), which is strong evidence that the two critical points belong to the same universality class, i.e., four-state Potts 
criticality.
 
Since the scaling collapse of the smallest gap, $\Delta$, is consistent with both Eq.~\eqref{eq:collapse} with $\nu \approx 0.72$ and 
Eq.~\eqref{eq:collapselog}, one may ask whether some other quantity can distinguish between the two scenarios. An analytical study using the
real-space 
renormalization group gives a relation between the critical exponents 
$\nu$ and the corresponding $\lambda$ of the AT model as~\cite{brien}
\begin{equation} \nu ~=~ \frac{1}{2 ~-~ (\frac{\pi}{2}) ~[\arccos(-\lambda)]^{-1}}, \end{equation}
which gives $\lambda=0.8267$ for $\nu=0.7233$ extracted from the behavior of $d(\Delta L)/dh$ at $h_c=1$ for various values of $L$ assuming a scaling of the form $L^{1/\nu}$(ignoring the statistical error in the fitting procedure). First, fitting $\Delta|_{h_c} \sim L^{-z}$ again yields $z \approx 0.97$ at this $\lambda$, similar to the 
case of the three-spin model and the AT model at $\lambda=1$. However, even though $\Delta|_{h_c}L$ seems to be consistent with Eq.~\eqref{eq:additivelog}, 
the extrapolated value of $a^* \approx 7.97$ turns out to be quite different from the corresponding value for the critical point of the three-spin 
model. Second, it turns out that probing Binder cumulants~\cite{binder,colin,wang}, denoted by $U_2$, constructed from the 
second and fourth moments of appropriately defined order parameters for the 
three-spin model, as well as the AT model for both $\lambda=1$ as well as $\lambda=0.8267$ helps to clarify the situations. 
$U_2$ can be defined such that it equals $1$ 
($0$) in the ordered (disordered) phase and attains a non-trivial value $U_2^*$ at the critical point in the thermodynamic limit. For finite systems, $U_2$ 
typically displays a monotonic behavior as a function of $h$, the parameter that drives the quantum phase transition, and thus stays bounded between $0$ 
and $1$. On the other hand, $U_2$ shows a non-monotonic behavior and in fact develops a negative peak whose location approaches the critical points and 
whose magnitude diverges polynomially with $L$ as $L \rightarrow \infty$\cite{capponi, vollmayr}. However, from the results for the 
classical Ashkin-Teller model in two dimensions\cite{jin}, $U_2$ is expected to display a {\it pseudo-first-order behavior} characterized by a non-monotonic behavior of 
$U_2$ in $h$ as well as a negative dip that increases in magnitude with system size, albeit much slower than what is expected from a first-order transition, 
for the one-dimensional quantum AT model in the neighborhood of $\lambda=1$ but not sufficiently away from it. Obtaining $U_2$ from DMRG for finite open chains 
for the three-spin model, and the AT model at $\lambda=1$ and $\lambda=0.8267$ respectively, one sees that while both the three-spin model and the AT model at 
$\lambda=1$ shows such pseudo-first-order behavior, $U_2$ displays a monotonic behavior and is bounded between $0$ and $1$ for the case of the AT model with 
$\lambda=0.8267$ for the available system sizes (see Appendix A for numerical plots and further discussion). We take this as evidence that the three-spin 
model, in fact, displays four-state Potts criticality and is not in the universality class of an intermediate critical point described by $\lambda 
\approx 0.83$.

Before ending the discussion of the DMRG results, we point out that we have carefully checked that our results are well converged with respect to the 
bond dimension. 
For all the calculations, apart from keeping track of energy convergence, we have also checked that the energy fluctuation in the obtained state $\langle H^{2} 
\rangle - \langle H \rangle^{2} \sim 10^{-10}$, ensuring that the state obtained at the end of the DMRG calculation is indeed an eigenstate of $H$.
While computing the energy gap, we have 
explicitly checked that the overlap between ground state and the first excited state obtained via DMRG is of the order $10^{-10}$, which ensures their
expected orthogonality.

\section{Presence of quantum many-body scars}
\label{scars}

\subsection{Non-integrability of the model}
\label{sec5a}

We now show that the three-spin model is different from the TFIM in that while the latter model is well-known to be integrable, the former seems to be
non-integrable. A common diagnostic to test integrability is to study the energy
level spacing statistics. In this section, we will study that level spacing for $H_3$ and discover that shows that the model is non-integrable. If the spectrum
of energies is sorted in increasing order so that $E_n$ is the $n$-th energy level, then we define the level spacing as \cite{atas, chavda}
\begin{equation} s_n = E_{n+1}-E_n. \end{equation}

The distribution of $s$, called $P(s)$, gives a way of testing the integrability
of the system. The system is integrable if $P(s)$ is Poisson-like and is 
non-integrable if $P(s)$ has a Wigner-Dyson distribution.
However, for many-body systems with a non-constant density of states, a new quantity proposed by Oganesyan and Huse~\cite{huse} is more useful and reliable. 
This quantity $\tilde{r}$ is defined as follows 
\begin{equation} \tilde{r}= \frac{\mathrm{min}(s_n, s_{n-1})}{\mathrm{max}(s_n, s_{n-1})}.
\label{tilder} \end{equation}

\begin{figure}[H]
\centering
\includegraphics[width=0.4\textwidth]{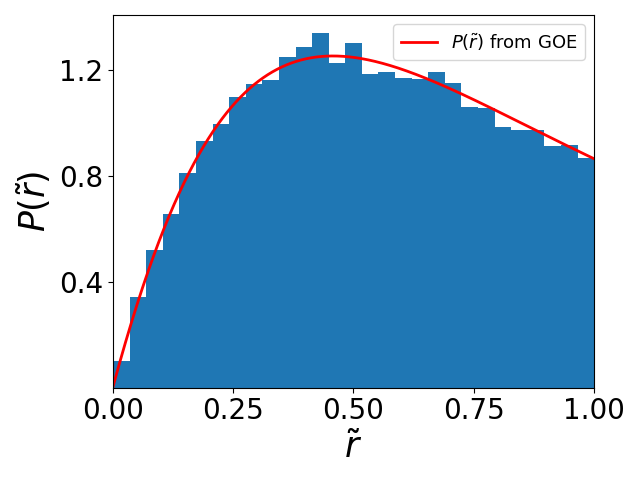}
\caption{The distribution of $\tilde{r}$ defined in Eq.~\eqref{tilder} is plotted for system size $L=18$ with open boundary conditions in the sector 
$(D_1, D_2, D_3) = (1,-1,-1)$ that contains $65536$ eigenstates. Further the expected distribution derived for a GOE $P(\tilde{r})$ is shown in red. We see 
that they agree quite well. The average 
of $\tilde{r}$ also turns out to be close to $0.53$ as expected for GOE.}
\label{fig:level_spacing}
\end{figure}

Since $\tilde{r}$ involves the ratio of energy spacings, the advantage of evaluating $\tilde{r}$ is that it is independent of the local density of states. 
The definition in Eq.~\eqref{tilder} implies that it is restricted to lie 
in the range $0$ to $1$.The average value of $\tilde{r}$ turns out to be $0.34$ 
for integrable models but close to $0.53$ for non-integrable models governed by
the Wigner-Dyson Gaussian orthogonal ensemble (GOE). For our model, we evaluate
$\tilde{r}$ in a particular sector $(D_1, D_2, D_3)= (1,-1,-1)$ and with open
boundary conditions to eliminate degeneracies due to any residual global symmetries. For $L=18$, we obtain the value of the average of $\tilde{r}$ to be $0.533$. We further see that the
numerical data fits very well to the Wigner-Dyson distribution of 
$P(\tilde{r})$ given by \cite{atas}
\begin{equation}
P(\tilde{r}) = \frac{27}{4} \frac{r+ r^2}{(1+ r+r^2)^{5/2}}\Theta(1-r),
\label{rtilde} \end{equation}
where $r_n= s_n/s_{n-1}$ and $\Theta(x)$ is the usual theta function. In Fig.~\ref{fig:level_spacing}, we see that the distribution given in 
Eq.~\eqref{rtilde} matches very well with the numerical data. This establishes that the three-spin model is non-integrable. Given that it is non-integrable, 
we find some of the energy eigenstates with zero energy have an
interesting feature 
as will be discussed in the section below. 

\begin{widetext}
\begin{figure*}
\centering
\includegraphics[width=0.8\textwidth]{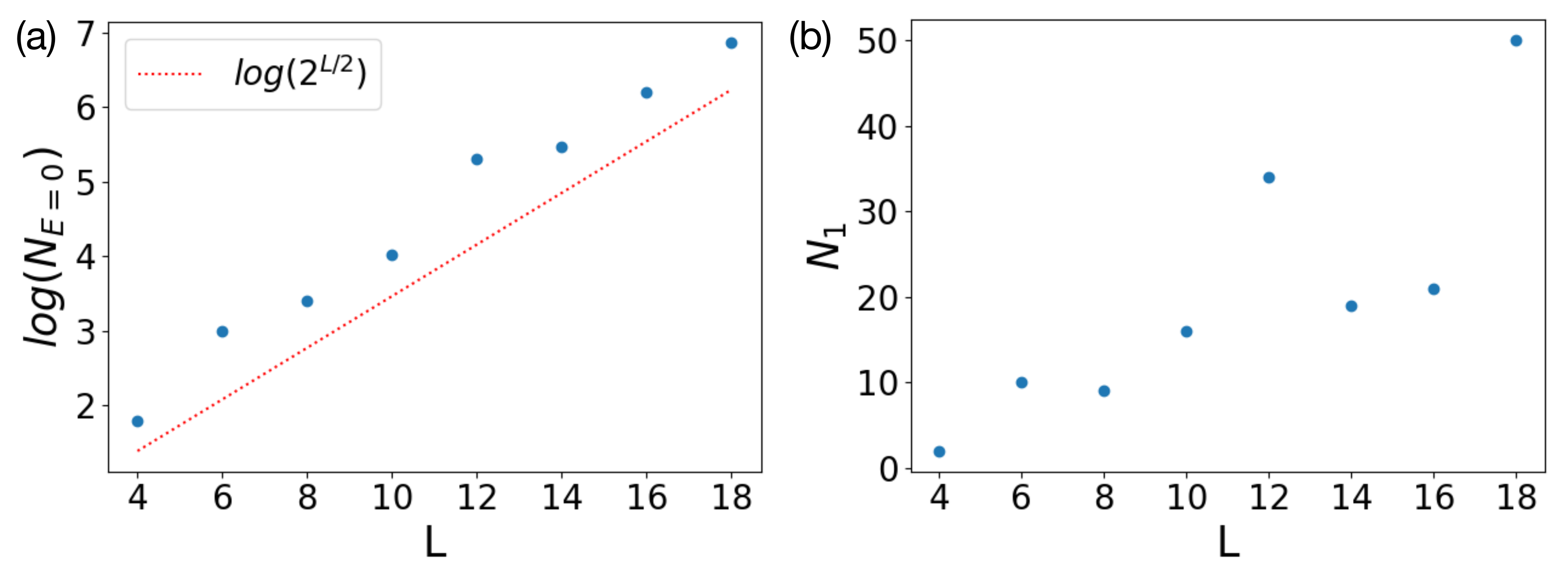}
\caption{(a) Plot of logarithm of total number of zero-energy states $N_{E=0}$ 
versus the system size $L$. For all $L$, we see that it is 
greater than $2^{L/2}$ which is a bound given by an index theorem. (b) Plot 
of the total number $N_1$ of Type-I states versus $L$. This number also
generally increases with system size although not monotonically.} 
\label{fig:zero_energy} \end{figure*}
\end{widetext}

\subsection{Zero-energy states}
\label{zeroenergy}

An the interesting property of the three-spin model is that for even system sizes with PBC, we find a large number of states with $E=0$. These are mid-spectrum 
states since 
we have a $E \rightarrow -E$ symmetry of the energy levels. We find that
the number of zero-energy increases with system size at least as fast as
$2^{L/2}$. We can prove this using an index theorem~\cite{michael2}. Writing the Hamiltonian $H_3$ in the $\sigma^y$ basis we find that it can be made to have 
only off-diagonal blocks when the states are divided into two sectors as follows. 
The spin states for a system size of $L$ is divided into sectors of (i) states with the number of spin states with $\sigma^y = +1$ being even, labeled as 
$N_{\uparrow, \text{even}}$ (ii) states with the number of spin states with $\sigma^y = +1$ being odd labeled as $N_{\uparrow, \text{odd}}$. This is 
because for states in $N_{\uparrow, \text{even}}$, the first term in $H_3$ 
will flip spins on three sites in the state and the second 
term will flip one spin, both giving a state with an odd number of up spins, thus connecting to the sector of $N_{\uparrow, \text{odd}}$. The index theorem 
states that the number of zero-energy states in the system is equal to or greater than the absolute value of the difference in the number of states in 
each sector, thus giving a lower bound on the number of zero-energy states. In this case, however, we find that $|N_{\uparrow, \text{even}}$- $N_{\uparrow, \text{odd}}| = 0$.
However we see that the parity operator can be used to further 
divide these two sectors into states with $P= \pm 1$. Since $L$ is even, we define parity as reflection about the middle of the $(\frac{L}{2})$-th and 
$(\frac{L}{2}+1)$-th sites and find the number of states with parity $P= \pm 1$ in the two sectors $N_{\uparrow, \text{even}}$ and $N_{\uparrow, 
\text{odd}}$. Let $n_1$ be the number of states with $(P= 1, N_{\uparrow, \text{even}})$, $n_2$ with $(P= 1, N_{\uparrow, \text{odd}})$, $n_3$ with $(P= 
-1, N_{\uparrow, \text{even}})$, and $n_4$ with $(P= -1, N_{\uparrow, \text{odd}})$. We know the following relations between $n_1$, $n_2$, $n_3$ 
and $n_4$. 
\begin{align*}
n_1 + n_2 +n_3 + n_4 &= 2^L, \\
n_1 + n_3 = n_2 + n_4& = 2^{L-1}.
\numberthis
\label{n1n2n3n4eqs}
\end{align*}

Next, given the spin configuration in one of the states, we can see that there are two possibilities. For the system size with $L$ sites, we can have the configuration from site numbers 1 to $L/2$ to be either (i) different from or (ii) same as the configuration from sites $L$ to $(L/2) +1$. Examples of this
for $L=6$ are as follows. For the first type, an example of such a configuration is a state like $\ket{\psi}$=
$\ket{\uparrow \uparrow \downarrow \uparrow \downarrow \uparrow}$ for which we see that the spins from $1$ to $3$ are not the same as from $6$ to $4$. For such states we can take superpositions $\ket{\psi} + P \ket{\psi}$ and $\ket{\psi} - P \ket{\psi}$ which are eigenstates of $P$ with eigenvalues $+1$ and $-1$ respectively. These two come in equal numbers for all $\ket{\psi}$. 
An example of the second type of configuration is $\ket{\uparrow \uparrow\downarrow \downarrow \uparrow \uparrow}$ 
where the reflection about the midpoint has the same configuration on either side. Such states therefore are eigenvectors of the parity operator with 
eigenvalue +1, i.e $P\ket{\psi}= \psi$. We also note that such states have to belong to the sector $N_{\uparrow, \text{even}}$ since the total number of up-
pointing spins is always twice the number of them till half the lattice. 
From this, we can conclude that the total number of such states of second type are 
equal to the difference between the number of $N_{\uparrow, \text{even}}$ with $P=1$ and $P=-1$. It is also equal to the number of ways of selecting the 
configuration from sites $1$ to $L/2$, since the other half is then fixed 
by mirror reflection. This gives $2^{L/2}$, which leads to the relation
\begin{equation} n_1 - n_3 = 2^{L/2}. \label{n1n3} \end{equation}
Turning to the sector $N_{\uparrow, \text{odd}}$, we see that no state can have 
$P\ket{\psi}= \pm \ket{\psi}$. The combination $\ket{\psi} + P \ket{\psi}$ and $\ket{\psi} - P \ket{\psi}$ again gives equal number of states with eigenvalues 
$\pm 1$. This further implies that
\begin{equation} n_2= n_4.
\label{n2n4} \end{equation}
From Eqs.~\eqref{n1n2n3n4eqs}, \eqref{n1n3} and \eqref{n2n4}, we have the following expressions for the numbers of states in the four sectors,
\begin{align*}
n_1 &= \frac{1}{2}(2^{L-1}+ 2^{L/2}), \\
n_2 &= \frac{1}{2} 2^{L-1}, \\
n_3 &= \frac{1}{2}(2^{L-1}- 2^{L/2}), \\
n_4 &= \frac{1}{2} 2^{L-1}.
\numberthis
\end{align*}
Thus considering the parity sector $P= +1$, we see that a lower bound for 
the number of 
zero-energy states is given by $|n_1 - n_2|= \frac{1}{2}2^{L/2}$, and similarly for $P= -1$, we have $|n_3- n_4|= \frac{1}{2}2^{L/2}$. Adding these up we see 
that the total number of zero-energy states for this system must satisfy
$N_{E=0} \geq 2^{L/2}$. We plot the total number of zero-energy states as a function 
of $L$ in Fig.~\ref{fig:zero_energy} (a). We indeed see that the number is 
greater than the lower bound of $2^{L/2}$ for all values of $L$.

\subsection{Type-I and Type-II zero modes}
\label{sec5c}

We now notice something more interesting about the zero-energy states described 
in Sec.~\ref{zeroenergy}. We again consider the Hamiltonian written in the
form given in Eq.~\eqref{hzx}. It then turns out that the zero-energy 
states come in two types, Type-I and Type-II. A given zero-energy state $\ket{\psi}$ is said to be Type-II if $H_3 \ket{\psi}=0 $ 
but the two terms separately do not give zero, i.e., $Z\ket{\psi} \neq 0$ and $X\ket{\psi} \neq 0$. However for a few of the 
zero-energy states, it turns out that the terms individually also give zero eigenvalues, that is, $Z\ket{\psi}=0$ and $X\ket{\psi}=0$. This means that the 
wave functions of these states are independent of the transverse field $h$.
These Type-I zero modes violate the ETH since they remain unchanged
as the coupling $h$ is varied in spite of the energy level spacing in their
neighborhood being exponentially small in $L$~\cite{deb} and can, therefore, be classified as quantum many-body scars~\cite{turner}. The number of these
Type-I zero-energy states $N_{1}$ also increases with system 
size as shown in Fig.~\ref{fig:zero_energy} (b). We do not know precisely
how fast $N_1$ grows with the system size $L$, but we will show below that
the growth is at least linear. The speciality of the Type-I states becomes more clear when we look at a plot of the half-chain entanglement entropy versus the
energy spectrum of this model. We find that most of the states lie close to the
thermal entropy of the system except for some states which stand out at $E=0$. 
These are the Type-I zero-energy states 
which turn out to typically have very low entanglement entropy compared to a generic state close to $E=0$ showing a violation of the ETH \cite{anza, luca}.
For a given system size, we can further perform a minimization of the
entanglement entropy within the subspace of these scar states \cite{sapta} using the algorithm outlined in Ref.~\onlinecite{Reuvers}. 
We show these plots with the full spectrum along with the entanglement-entropy minimized scar states in Fig.~\ref{fig:ent1} (a) and (b), for system sizes 
$L=12$ and $L=18$. We see that there is a dramatic drop in the entropy for 
most of these scar states confirming that they indeed violate the ETH.

The total number of zero-energy states and the number of Type-I zero-energy states
for various system sizes $L$ are shown in Table~\ref{table2}. We see that the total
number of zero-energy states increases rapidly with $L$ while the number of
Type-I states changes non-monotonically but on the average increases with $L$.

We can further appreciate the difference between Type-II and Type-I states by studying their distribution over the Fock space. A state can be written as a 
superposition of the basis states of the entire Fock space. A particular scar state, after it has been minimized for entanglement entropy, can be written as 
$\ket{\psi_{S}} = \sum_n^{2^L} c_n \ket{\psi_n}$, where $\ket{\psi_n}$ are the basis states in the Fock space and $c_n$ is the corresponding amplitude for the 
scar state $\ket{\psi_{S}}$. In Fig.~\ref{fig:scardist}, we plot the probability $|c_n|^2$ for a generic zero-energy state and for the scar states. 
We see that a Type-II state (Fig.~\ref{fig:scardist} (a)) has non-zero coefficients over a large number of basis states, and the distribution looks random. However, 
Type-I states as shown in Figs.~\ref{fig:scardist} (b) and (c) can be easily distinguished as they have a large weight over only a few basis states with 
equal probabilities.
\vspace*{.2cm}

\begin{table}[H]
\begin{center}
\begin{tabular}{||c| c| c||} 
\hline
System size & Total number of & Number of \\ [0.5ex] 
$L$ & zero-energy states & Type-I states \\ [0.5ex] 
\hline
 $4$ & 6 & 2 \\ [1ex] 
 \hline
 $6$ & 20 & 10 \\ [1ex] 
 \hline
 $8$ & 30 & 9 \\ [1ex] 
\hline
 $10$ & 56 & 16 \\ [1ex] 
\hline 
$12$ & 202 & 34 \\ [1ex] 
\hline
$14$ & 236 & 19 \\ [1ex] 
\hline
$16$ & 492 & 21 \\ [1ex] 
\hline 
$18$ & 970 & 50 \\ [1ex] 
\hline
\end{tabular}
\end{center}
\caption{Total number of zero-energy states and number of Type-I zero-energy
states for various system sizes.} \label{table2}
\end{table}

An interesting feature of the Type-I states is that since they are annihilated
simultaneously by the operators $X$ and $Z$ and are therefore zero 
energy eigenstates of the Hamiltonian $H_3$ for any value of $h$,
they will not evolve with time even if $h$ varies with time in an arbitrary
way. In particular, if $h$ is taken to vary periodically with a time
period $T$, the Type-I states will be eigenstates of the Floquet operator
$U_T$ which evolves the system over one time period, and the eigenvalue will
be exactly equal to 1. While examples of quantum many-body scars has been found 
in driven systems for specific driving protocols~\cite{bhaskardrive1, Sugiuradrive,bhaskardrive2, Mizutadrive,bhaskardrive3,bhaskardrive4}, 
these Type-I states provide examples of scars for any driving protocol $h(t)$.
Similarly, if the value of $h$ is suddenly changed from one value to
another (called a quench), these states will not change. Finally, 
if $h$ is held fixed and we initialize the system in one of these states 
(or in an arbitrary linear combination of them), the system will stay in 
that state for all times, namely, the system will not thermalize.

\begin{widetext}
\begin{figure*}
\centering
\includegraphics[width=0.8\textwidth]{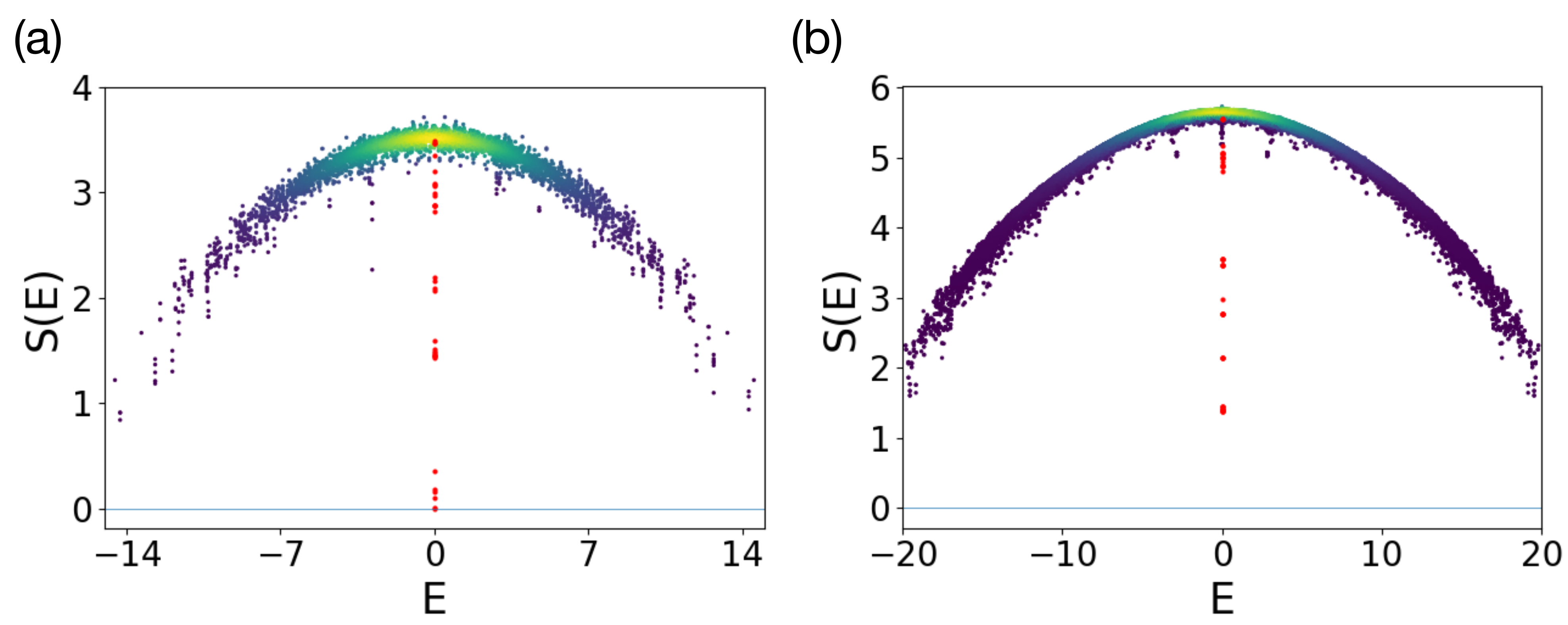}
\caption{Plots of the half-chain entanglement entropy spectrum for all the energy 
levels of the system for $L=12$ and $L=18$ are shown in (a) and (b) respectively. 
The plot in red correspond to the Type-I scar states which have entanglement 
entropy much lower than the neighbouring states, clearly violating the ETH.} 
\label{fig:ent1} \end{figure*} 

\begin{figure*}
\centering
\includegraphics[width=1.0\textwidth]{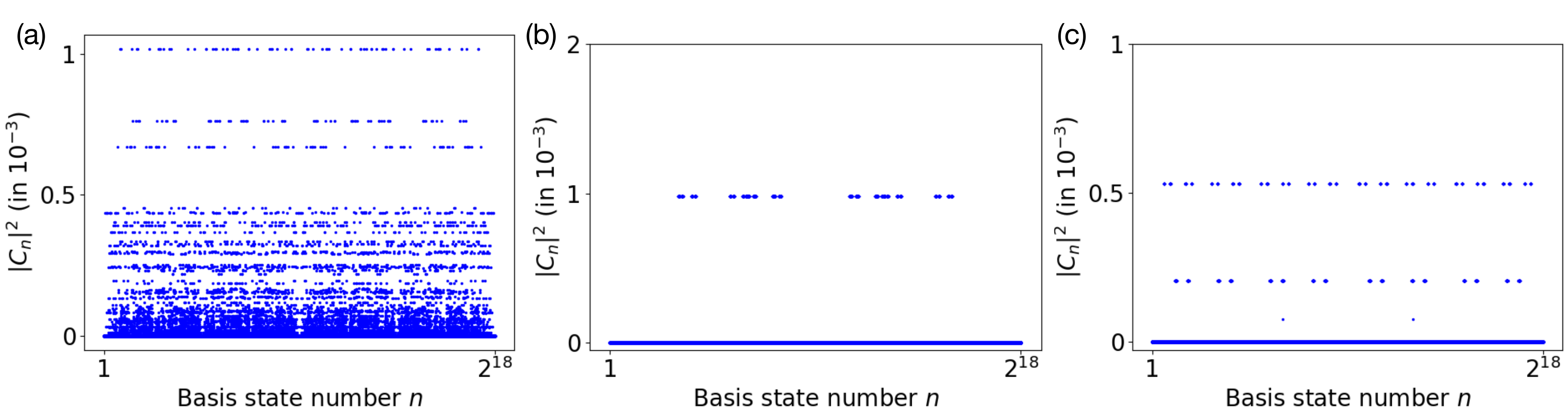}
\caption{(a) Probabilities $|c_n|^2$ of a generic Type-II $E=0$ state in the entire Fock space for all the basis states are plotted for $L=18$. We see that the distribution is random. (b) and (c) show the same plot for different Type-I scar states. The distribution is more sparse and also has equal probabilities 
for many basis states.} \label{fig:scardist} \end{figure*}
\end{widetext}

\subsection{Some exact Type-I states}
\label{sec5d}

\begin{figure}
\centering
\includegraphics[width=0.45\textwidth]{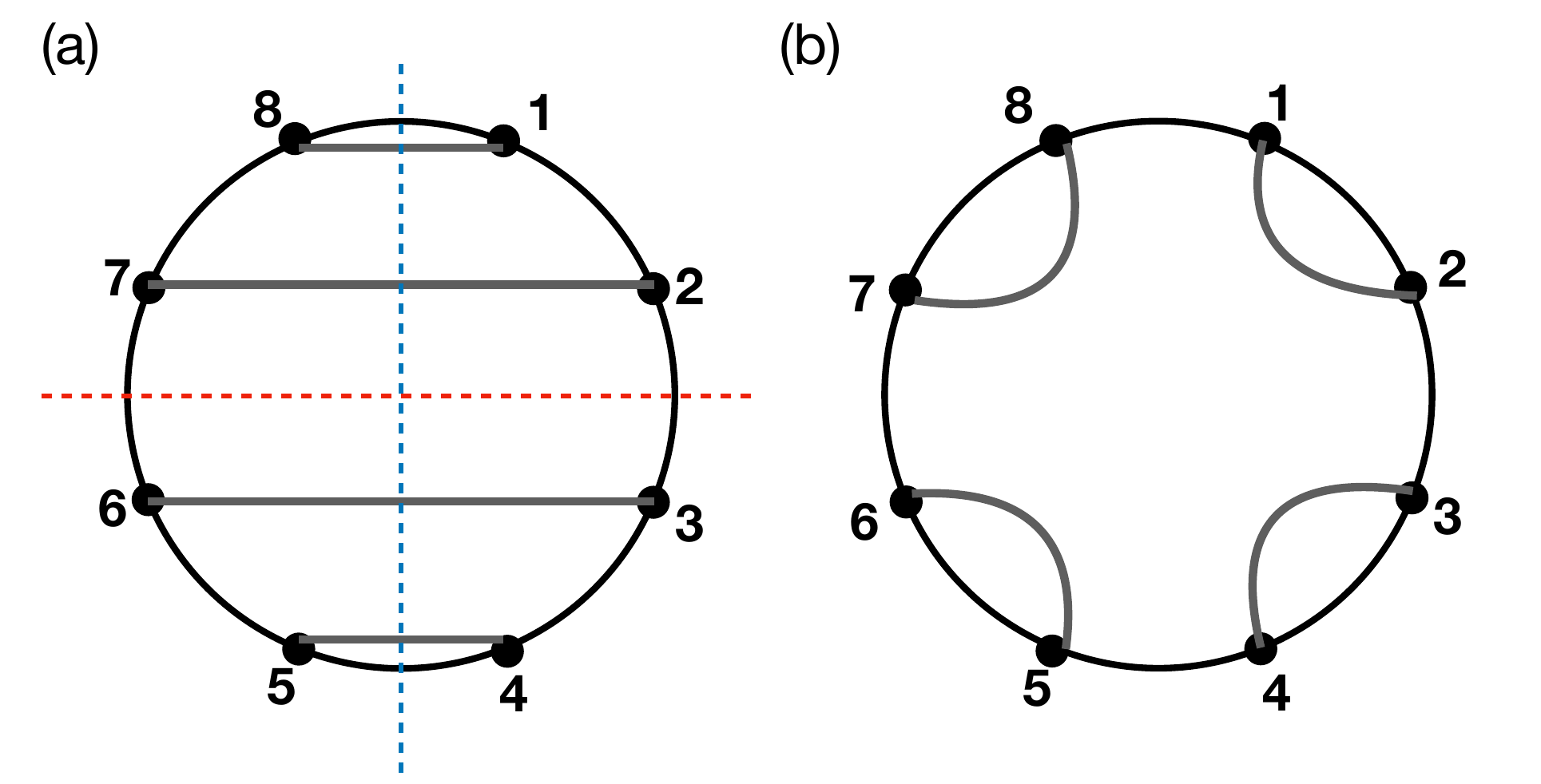}
\caption{(a) Picture of the state $| \psi_1 \ra$ given in Eq.~\eqref{psi1} for
$L=8$. The lines joining pairs of sites denote spin singlets. Two straight 
lines dividing the system into two equal parts are shown by dashed lines. The
vertical dashed line cuts $L/2$ singlets, while the horizontal dashed line does
not cut any singlet; thereby producing half-chain entanglement entropies
equal to $(L/2) \ln 2$ and zero respectively. (b) Picture of the state $| 
\phi_1 \ra$ given in Eq.~\eqref{phi12}.} \label{fig:typeI} \end{figure}

We will now present some Type-I states (scars) which we have found 
analytically~\cite{hosho}.
To this end, let us define two states involving sites $j$ and $k$ given by
\bea S_{j,k} &=& \frac{1}{\sqrt{2}} ~(| \ua_j \da_k \ra ~-~ | \da_j \ua_k \ra), 
\non \\
T_{j,k} &=& \frac{1}{\sqrt{2}} ~(| \ua_j \da_k \ra ~+~ | \da_j \ua_k \ra), 
\label{stjk} \eea
where $\ua$ and $\da$ denote spin-up and spin-down in the $\si^x$ basis.
These are, respectively, spin-singlet and spin-triplet states with total
$S^x = (\si_j^x + \si_k^x)/2 = 0$. Note that these states are antisymmetric
and symmetric respectively under the exchange of sites $j$ and $k$. We find
that they satisfy the identities
\bea \si_j^z ~S_{j,k} &=& - ~\si_k^z ~S_{j,k}, ~~~~ \si_j^z \si_k^z ~S_{j,k} 
~=~ - ~ S_{j,k}, \non \\
\si_j^z ~T_{j,k} &=& \si_k^z ~T_{j,k}, ~~~~~~~~~\si_j^z \si_k^z ~T_{j,k} 
~=~ T_{j,k}. \label{stjk2} \eea
We now consider a system with $L$ sites with PBC and a state which is a product
of singlets with the form
\beq | \psi_1 \ra ~=~ S_{L,1} ~S_{L-1,2} ~S_{L-2,3} ~\cdots S_{(L/2)+1,L/2}.
\label{psi1} \eeq
Clearly $X | \psi \ra = 0$ where the operator $X$ is given in Eq.~\eqref{hzx}.
(A picture of $| \psi_1 \ra$ for $L=8$ is shown in Fig.~\ref{fig:typeI} (a)). 
Each line connecting a pair of sites denotes a spin-singlet state).
Eqs.~\eqref{stjk2} then imply that 
\bea (\si_{L-1}^z \si_L^z \si_1^z + \si_L^z \si_1^z \si_2^z) ~| \psi_1 \ra 
&=& 0, \non \\
(\si_{L-2}^z \si_{L-1}^z \si_L^z + \si_1^z \si_2^z \si_3^z) ~| \psi_1 \ra 
&=& 0, \non \\
(\si_{L-3}^z \si_{L-2}^z \si_{L-1}^z + \si_2^z \si_3^z \si_4^z) ~| \psi_1 \ra 
&=& 0, \non \\
\cdots && \non \\
(\si_{L/2+1}^z \si_{L/2+2}^z \si_{L/2+3}^z + \si_{L/2-2}^z \si_{L/2-1}^z 
\si_{L/2}^z) ~| \psi_1 \ra &=& 0, \non \\
(\si_{L/2}^z \si_{L/2+1}^z \si_{L/2+2}^z + \si_{L/2-1}^z \si_{L/2}^z 
\si_{L/2+1}^x) ~| \psi_1 \ra &=& 0. \non \\
&& \label{sumstjk} \eea
Hence the state $| \psi \ra$ satisfies $Z | \psi \ra = 0$ where
the operator $Z$ is given in Eq.~\eqref{hzx}. Since both $X$ and $Z$ 
annihilate $| \psi \ra$, we conclude that this is a Type-I state.

Next, we can take the state $|\psi_1 \ra$ and rotate all the sites clockwise
by 1 site on the circle. This gives the state 
\beq | \psi_2 \ra ~=~ S_{1,2} ~S_{L,3} ~S_{L-1,4} ~\cdots~ S_{(L/2)+2,(L/2)+1},
\label{psi2} \eeq
and following similar arguments we can show that $| \psi_2 \ra$ is also a 
Type-I state. Continuing in this way, we find $L/2$ distinct states, denoted
$| \psi_n \ra$, $n=1,2,\cdots,L/2$, which are all Type-I states. 

Now we observe that if the system is cut into two equal parts by a line, and we
consider the state $| \psi_1 \ra$, the line may cut no singlets, one singlet, 
two singlets, and so on all the way up to $L/2$ singlets, depending on the 
orientation of the line (see the two dashed lines in Fig.~\ref{fig:typeI} (a)). 
As a result,
the half-chain entanglement entropy can take all possible values from zero up 
to $(L/2) \ln 2$. Even the largest of these values is only half of the thermal
entropy given by $L \ln 2$. This again confirms that these are all scar states.
We note that the state shown in Fig.~\ref{fig:typeI} (a) resembles the
rainbow scars discussed in Ref.~\onlinecite{langlett}.

It turns out that there are two other singlet states, denoted $| \phi_1 \ra$
and $| \phi_2 \ra$, which are also Type-I states. These have the form
\bea | \phi_1 \ra &=& S_{1,2} ~S_{3,4}~ S_{5,6} ~\cdots~ S_{L-1,L}, \non \\
| \phi_2 \ra &=& S_{2,3} ~S_{4,5}~ S_{6,7} ~\cdots~ S_{L,1}. \label{phi12}
\eea
(A picture of $| \phi_1 \ra$ is shown in Fig.~\ref{fig:typeI} (b)).
Using Eqs.~\eqref{stjk2} we can show that these states are also annihilated
by the operator $Z$. (As before, we can find pairs of three-spin terms
$\si_i^z \si_j^z \si_k^z$ and $\si_l^z \si_m^z \si_n^z$ such that the sum
of the two terms annihilates the states $| \phi_n \ra$). Further, the half-chain 
entanglement entropy for these two states range from zero to $2 \ln 2$ depending 
on the orientation of the line which cuts the system into two halves.

For $L=4$, the states $\psi_n$ and $\phi_n$ are identical, and we therefore
have only two exact type-I states; according to Table II, these form the
complete set of Type-I states. For $L \ge 6$, the states $\psi_n$ and
$\phi_n$ are distinct, and we therefore have $(L/2) + 2$ Type-I states.

The states $| \psi_n \ra$ and $| \phi_n \ra$ discussed above are examples of
resonating valence bond (RVB) states for a $L$-site system. If the $L$ sites 
are arranged around a circle, the RVB states 
correspond to joining pairs of sites by lines in such a way
that no two lines cross each other. According to the Rumer-Pauling rules~\cite{pauling}, there are $L! /(L/2)! ((L/2)+1)!$ such states 
which are linearly independent, although not orthogonal to each other. We see
that $(L/2)+2$ of the RVB states are Type-I states for our model.
We conclude that the number of Type-I states increases at least linearly
with $L$.

\begin{figure}
\centering
\includegraphics[width=0.45\textwidth]{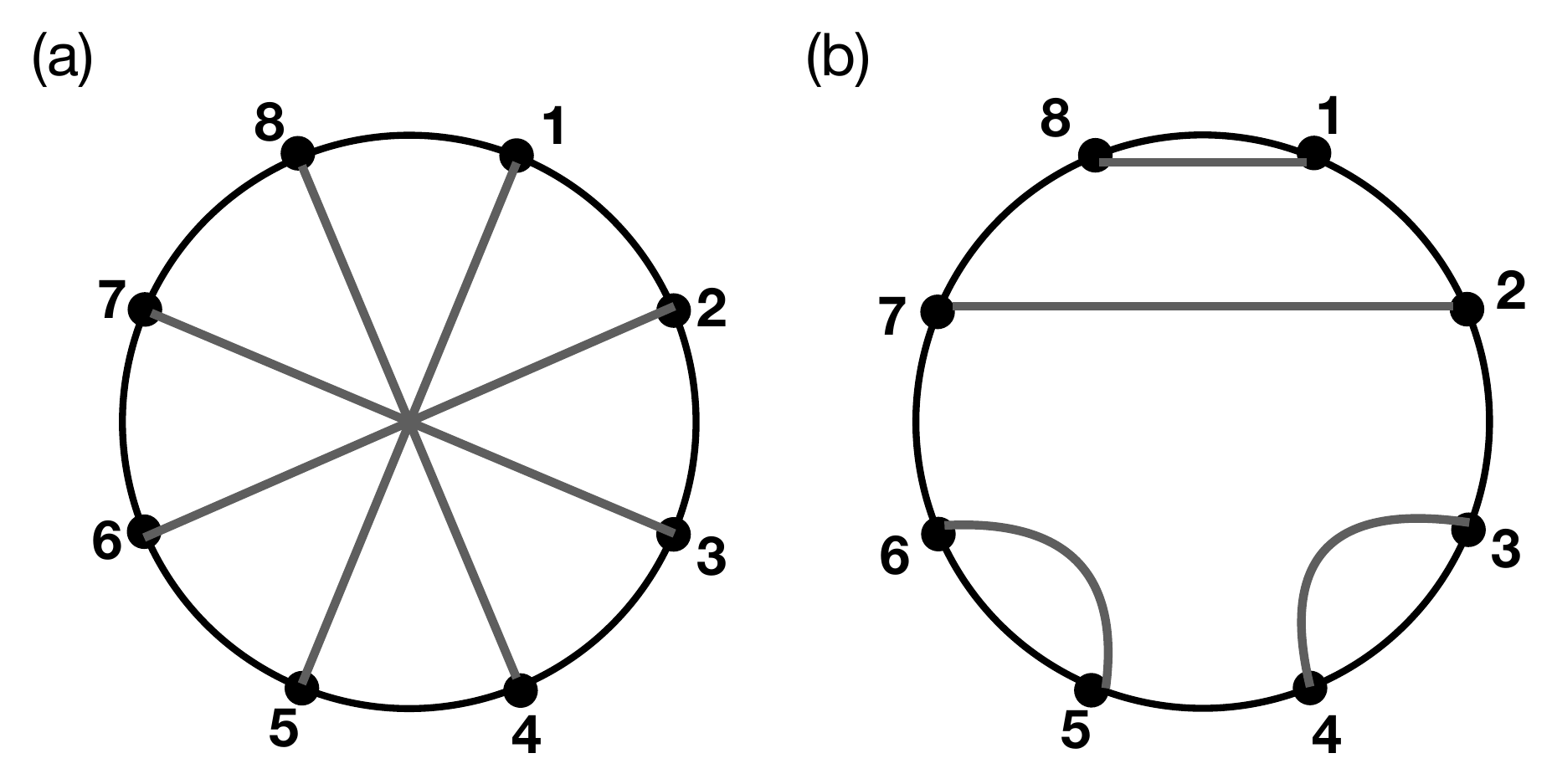}
\caption{(a) Picture of the state $| \psi_d \ra$ given in Eq.~\eqref{psi13a} 
for $L=8$. The lines joining pairs of diametrically opposite sites denote spin
singlets. (b) Picture of the state $|\phi_d \ra$ given in Eq.~\eqref{psi13b}.}
\label{fig:typeIb} \end{figure}

We can construct one more Type-I state using singlet states as follows.
For a system with $L$ sites and PBC, consider the following state which is
a product of singlets connecting diametrically opposite sites,
\beq | \psi_d \ra ~=~ S_{1,(L/2)+1} ~S_{2,(L/2)+2} ~S_{3,(L/2)+3} ~\cdots~ S_{L/2,L}. 
\label{psi13a} \eeq
We find that this state is annihilated by terms of the form
$\si_n^z \si_{n+1}^z \si_{n+2}^z + \si_{(L/2)+n}^z \si_{(L/2)+n+1}^z 
\si_{(L/2)+n+2}^z$,
where $n=1,2,\cdots,L/2$. Hence $| \psi_d \ra$ is annihilated by the operator $Z$
given in Eq.~\eqref{hzx}. A picture of $| \psi_d \ra$ for $L=8$ is shown in 
Fig.~\ref{fig:typeIb} (a). However, $| \psi_d \ra$ is not an RVB state since the
different singlet lines cross each other; in fact, any two singlet lines cross
each other. But we can write $| \psi_d \ra$ as a linear combination
of RVB states by using the identity
\beq S_{i,j} ~S_{k,l} ~-~ S_{i,k} ~S_{j,l} ~+~ S_{i,l} ~S_{j,k} ~=~ 0
\label{iden} \eeq
several times.
Depending on how four sites labeled $i, ~j, ~k, ~l$ are arranged around a circle,
one of the terms in Eq.~\eqref{iden} will corresponding to a state with one
crossing while the other two terms will correspond to non-crossing states.
Hence, by repeatedly using Eq.~\eqref{iden}, we can successively decrease the 
number of crossings to eventually reduce $| \psi_d \ra$ to a superposition of 
RVB states. For $L=8$, we find that the superposition contains all the states
shown in Figs.~\ref{fig:typeI} (a) and (b) as well as a {\it specific} linear
combination of 8 other RVB states which are of the form 
\beq | \phi_d \ra ~=~ S_{1,8} ~S_{2,7} ~S_{3,4} ~S_{5,6}, \label{psi13b} \eeq
shown in Fig.~\ref{fig:typeIb} (b), and 7 other states obtained from Eq.~\eqref{psi13b} by rotating all the sites clockwise by $1, ~2, ~\cdots, ~7$ 
sites.

The different kinds of exact Type-I states discussed above do not exhaust 
all the Type-I states. For instance, Table II shows that there are 9 Type-I 
states for $L=8$, but the arguments above only account for $(L/2)+2+1 = 7$ of them.

\begin{figure}
\centering
\includegraphics[width=0.45\textwidth]{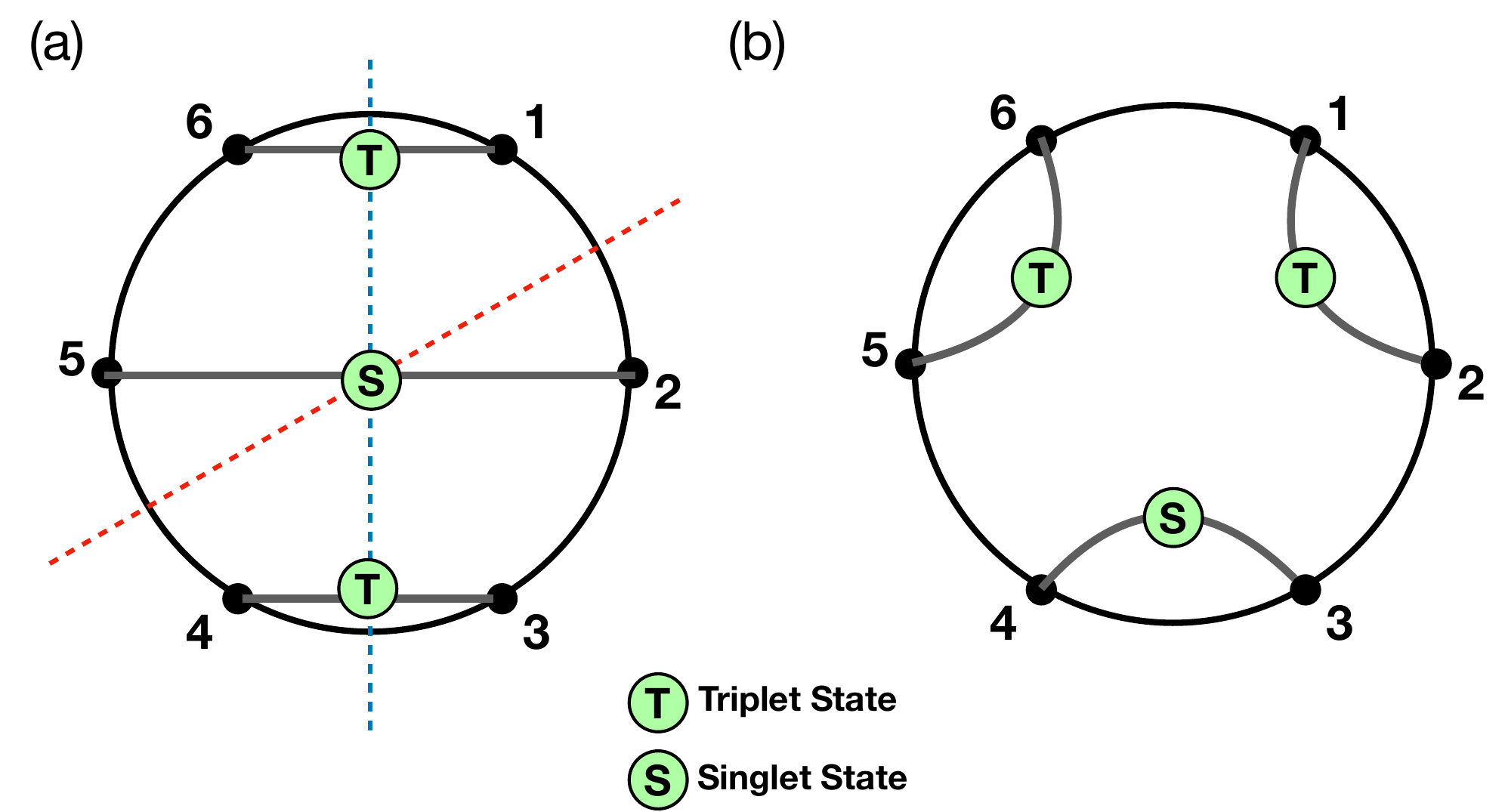}
\caption{(a) Picture of the state $| \psi'_1 \ra$ given in Eq.~\eqref{psip1} 
for $L=6$. The lines joining pairs of sites denote spin singlets or triplets
shown as $S$ or $T$ respectively. The dashed lines divide the system
into two equal parts and they cut one and three bonds respectively. 
(b) Picture of the state $| \phi'_1 \ra$ given in Eq.~\eqref{phip1}.}
\label{fig:typeIc} \end{figure}

Finally, we note that if $L$ is a multiple of 6, we can find exact Type-I states
involving both singlets and triplets. Two such states are shown in
Figs.~\ref{fig:typeIc} for a system with 6 sites. The state in 
Fig.~\ref{fig:typeIc} (a) has the form
\beq | \psi'_1 \ra ~=~ T_{6,1} ~S_{5,2}~ T_{4,3}, \label{psip1} \eeq
while the state in Fig.~\ref{fig:typeIc} (b) has the form
\beq | \phi'_1 \ra ~=~ T_{1,2} ~S_{3,4}~ T_{5,6}. \label{phip1} \eeq
These can be shown to be Type-I states by similar arguments as above and using the
identities in Eqs.~\eqref{stjk2}. Then one can repeatedly rotate all the sites 
by 1 site from $|\psi'_1 \ra$ and $| \phi'_1 \ra$ obtain states of the form 
$| \psi'_n \ra$, $n=1,2,3$, and $| \phi'_n \ra$, $n=1,2,\cdots,6$, respectively.
We thus obtain 9 states each of which involves one singlet and two triplets.
However, one can show that only 5 of these states are linearly independent;
one can choose these to be of the form $| \phi'_n \ra$, where 
$n=1,2,\cdots,5$. 
For $L=6$, therefore, we get $(L/2) +2 = 5$ states involving only singlet states
and 5 state involving both singlets and triplets. This gives a total of 
$10$ Type-I states for $L=6$ which is in agreement with Table II. 

A similar construction of Type-I states involving singlets and triplets exists
whenever $L$ is a multiple of 6. There are two kinds of such states.
The first kind of states resembles the one shown in Fig.~\ref{fig:typeIc} (a) 
and is given by
\bea | \psi'_1 \ra &=& T_{L,1} ~S_{L-1,2} ~T_{L-2,3} ~T_{L-3,4} ~\cdots~
S_{L-4,5} \non \\ 
&& \cdots ~T_{L/2+3,L/2-2} ~S_{L/2+2,L/2-1} ~T_{L/2+1,L/2}, \non \\
&& \label{psip2} \eea
and similar states obtained by rotating all the sites by 1 site. The pattern
of bonds from the top to the bottom follows the pattern $TSTTSTT \cdots STTST$. 
There are $L/2$ states of this kind. The second kind of states resembles the 
one shown in Fig.~\ref{fig:typeIc} (b) and is given by
\bea | \phi'_1 \ra &=& T_{1,2} ~S_{3,4}~ T_{5,6} ~T_{7,8} ~S_{9,10} ~T_{11,12}
\non \\
&& \cdots~ T_{L-5,L-4} ~S_{L-3,L-2} ~T_{L-1,L}, \label{phip2} \eea
and similar states obtained by rotating all the sites by 1 site. The bonds
follow the pattern $TSTTSTT \cdots STTST$ around the circle. There are $6$ states
of this kind. We therefore have a total of $(L/2) + 6$ Type-I states, each of which 
is a product of $L/6$ singlets and $L/3$ singlets. However, unlike the 
Type-I states which involve only singlets where we found that there are $(L/2) + 2$
linearly independent states, we do not know how many of the $(L/2) + 6$ Type-I states
involving singlets and triplets are linearly independent for a general value of
$L$. For $L=6$, we saw above that the number of independent states is 5, but a 
formula for $L=12,18,24,\cdots$ is not known.

Once again, we note that when the system is cut into two equal parts by 
a line, the number of bonds (singlets or triplets) cut by the line can vary
from zero to $L/2$ depending on its orientation (see 
Fig.~\ref{fig:typeIc} (a)). The half-chain entanglement entropy can therefore
vary from zero up to $(L/2) \ln 2$ which is much smaller than the thermal
entropy equal to $L \ln 2$.

The fact that there are several exact Type-I states involving singlets and 
triplets which appear only when $L$ is a multiple of 6 may help to explain 
why there is a jump in the number of Type-I states whenever $L$ hits those
particular values, as we can see in Table II.

It would be interesting to find all possible Type-I states exactly but this 
seems to be a 
difficult problem. We note that all the exact Type-I states discussed in this 
section have been found by demanding that they be annihilated by the sum of 
two three-spin terms of the form $\si_i^z \si_j^z \si_k^z + \si_l^z \si_m^z 
\si_n^z$, and these sums combine to give the total operator $Z$ in Eq.~\eqref{hzx}.
However, there may be more complicated Type-I states
which are only annihilated by the sum of three or more three-spin terms.

It is intriguing that singlets and triplets (with zero magnetization)
play such an important role in the construction of Type-I states even though the Hamiltonian $H_3$ is not invariant under $SU(2)$ or any other continuous symmetry.

\section{Anomalous relaxation of autocorrelators at different sites}
\label{correlator}

The ordered phase of the TFIM on a semi-infinite system is characterized by a doubly degenerate spectrum and the presence of a strong edge mode operator that 
connects pairs of degenerate states with opposite parity~\cite{fendley2012, fendley2016}. Numerically, this can be observed by studying the 
infinite-temperature autocorrelator of the $\sigma^{z}$ operator at different sites near the edge of the system\cite{ kemp2017,yates2020,yates2020prb}. 
\begin{equation} A^{zz}_{l}(t) = \frac{1}{2^{L}} \text{Tr}[\sigma_{l}^{z}(t)
\sigma_{l}^{z}]. \end{equation}
Since the strong mode operator has a large overlap with the operator $\sigma^{z}_1$ operator at the boundary site, the autocorrelator shows a long
plateau near the value of unity with a time scale that increases exponentially
with the system size before relaxing to zero. However the autocorrelator of $\sigma^{z}$ at any other site falls off to zero 
very quickly in a time scale $t \lesssim 10 $. 

\begin{figure*}
\centering
\includegraphics[width=\textwidth]{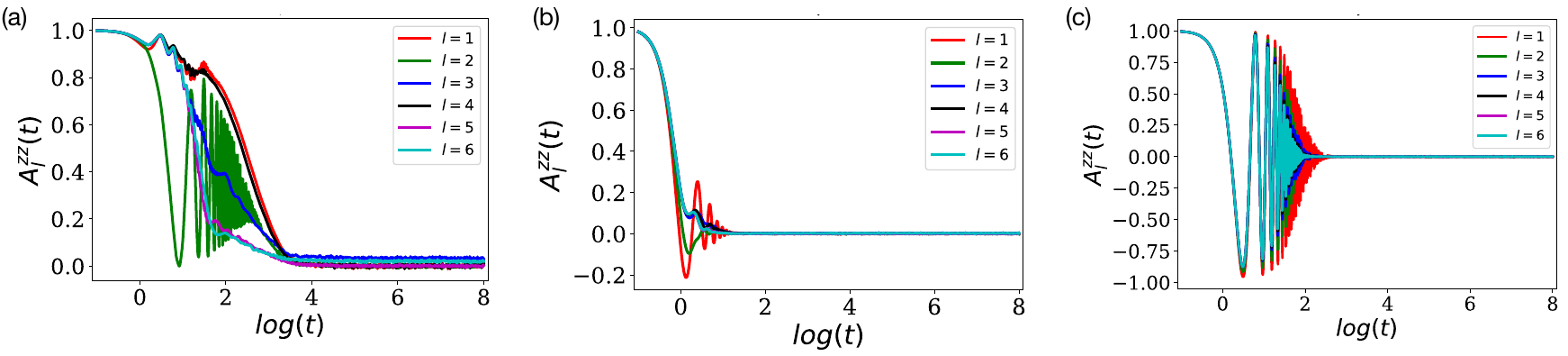}
\caption{Autocorrelation function $A_l^{zz} (t)$ for $L=14$ plotted versus time on 
a log scale for different values of the transverse field. Deep inside the
ordered phase, (a) $h = 0.2$, or the disordered phase, (c) $h=5$, the
autocorrelators at several sites near the boundary show oscillations for a long
time before decaying to zero. (b) At the critical point, $h=1$, the autocorrelators
decay quickly to zero at all sites except at the boundary site.}
\label{fig:sz_autocorr1} \end{figure*}

This motivates us to ask a similar question for the non-integrable model
$H_3$ with open boundary conditions. As discussed earlier, this model has an exact degeneracy in three-fourths of its eigenstates due to the presence of the $D_1, ~D_2, ~D_3$ operators for PBC, and also a 
two-fold degeneracy in half of its eigenstates due to parity symmetry for
open boundary conditions. These degeneracies are present for any value of the transverse field $h$. We will study how the spin autocorrelators relax in time
at sites near the boundary for various values of $h$ and see if the degeneracies
play any role in the relaxation. The infinite-temperature autocorrelators can be
calculated as traces over all the energy eigenstates of the Hamiltonian. We will
be interested in the $zz$- and $xx$-autocorrelators given by
\begin{equation} A^{zz}_{l}(t) = \frac{1}{2^{L}} \sum_{n,m} e^{~i (E_n -E_m)t}
|\langle n |\sigma_{l}^{z} |m \rangle|^{2}, \label{Auto_statesz} \end{equation}
and
\begin{equation} A^{xx}_{l}(t) = \frac{1}{2^{L}} \sum_{n,m} e^{~i (E_n -E_m)t} |\langle n |\sigma_{l}^{x} |m \rangle|^{2}, \label{Auto_statesx} \end{equation}
respectively.
The autocorrelators defined in this way are expected to reveal the nature of the phase transition and the energy spectra on the two sides of the transition.

\begin{figure*}
\centering
\includegraphics[width=0.8\textwidth]{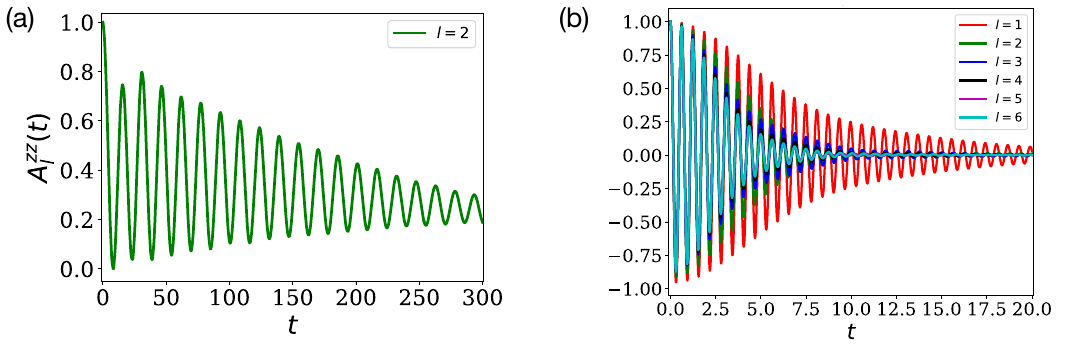}
\caption{(a) $A^{zz}_{l}(t)$ at site $l=2$ for $h=0.2$, showing long-time
oscillations. This can be understood using first-order degenerate perturbation
theory. (b) $A^{zz}_{l}(t)$ showing oscillations at different sites for $h=5$. 
This can be understood using effective two-level systems. Both the figures are for system size $L=14$.} \label{fig:sz_autocorr2} \end{figure*}

We present the results for $A^{zz}_{l}(t)$ versus $t$ on a log scale for
different lattice sites $l = 1,2, \dots 6$ (with $l=1$ being the boundary site) and three values of the transverse field, $h = 0.2, 1,$ and $5.0$, in Figs.~\ref{fig:sz_autocorr1} (a), (b) and (c) respectively. The relaxation of 
the autocorrelators shows very interesting behaviors depending on whether 
$h \ll 1$, $h \gg 1$ or $h=1$. For $h = 0.2$ (see Fig.~\ref{fig:sz_autocorr1} (a)), we observe qualitatively that $A^{zz}_{1}$ and $A^{zz}_{4}$ have a similar
structure, with a small plateau for a time interval of $t \lesssim 10^{4}$, where the autocorrelator remains near 1 before falling off to zero at large times.
We believe that this is due to the presence of an operator, which has an 
appreciable overlap with $\sigma^{z}$ at sites $1$ and $4$ and also has a small commutator with the Hamiltonian itself. The autocorrelator at site $l = 2$ 
has the most striking behavior, showing oscillations with an approximate period 
of $15.5$. We also plot the same autocorrelator in real time instead of the 
logarithmic scale in Fig.~\ref{fig:sz_autocorr2} (a), where the oscillations 
can be seen clearly. 

The small frequency oscillations at the site $l=2$ can be explained by considering the Hamiltonian in the small $h$ limit and doing a perturbative
calculation. First, by putting $h = 0$, we have the Hamiltonian given by $H_3 \Big|_{h = 0} = Z = -\sum_{j=1}^{L-2} \sigma^{z}_{j} \sigma^{z}_{j+1} 
\sigma^{z}_{j+2}$. The eigenstates of this are given by product states where each site $j$ has a definite value of $\sigma^{z}_j = \pm 1$. Therefore, all the 
eigenvalues of $Z$ are integer valued and so are the energy differences. Now,
an introduction of a small value of transverse field $h$ gives eigenstates 
with energy differences of order $h$. To see this, we look at the couplings in 
the $Z$ term and the effects of $\sigma^{z}_l$ in the autocorrelator more 
carefully. The 
couplings in $Z$ containing a particular $\sigma^{z}_{l}$ can be considered for three separate cases, (a) $\sigma^{z}_{l} (\sigma^{z}_{2} \sigma^{z}_{3})$, for 
$l = 1$, (b) $\sigma^{z}_{l} (\sigma^{z}_{1} \sigma^{z}_{3} + \sigma^{z}_{3} \sigma^{z}_{4})$, for $l = 2$, and (c) $\sigma^{z}_{l} (\sigma^{z}_{l-2} 
\sigma^{z}_{l-1} + \sigma^{z}_{l-1} \sigma^{z}_{l+1}+ \sigma^{z}_{l+1} \sigma^{z}_{l+2})$, for $l \ge 3$. Since each $\sigma^z_{j}$ can take values 
$\pm 1$, the products of two spin operators also will take values $\pm 1$. Therefore, in cases (a) and (c), we have a sum of an odd number of such products 
which necessarily has a non-zero value. However, in case (b), 
we have an even number of such terms and hence, for $l=2$, we can have a case where $\sigma^{z}_{2}$ is multiplied by zero. More precisely, this happens if 
$\sigma^{z}_{3}(\sigma^{z}_{1} + \sigma^{z}_{4})= 0$, i.e., if $(\sigma^{z}_{1} +
\sigma^{z}_{4}) = 0$. Thus, the two sets of eigenstates of $Z$ 
corresponding to the value of $\sigma^z_{2} = \pm 1$ (we label them as $\ket{I}$ and $\ket{II}$ respectively) will be degenerate for any values of 
$\sigma^{z}_{1}, \sigma^{z}_{3}, \sigma^{z}_{4}, \sigma^{z}_{5}, \sigma^{z}_{6}, \dots $, with the condition that $(\sigma^{z}_{1} + \sigma^{z}_{4}) = 0$. This 
condition is satisfied for half of the states when $\sigma^{z}_{1}$ and $\sigma^{z}_{4}$ are opposite to each other, and then we have a pairwise 
degeneracy between the states of types $\ket{I}$ and $\ket{II}$. With a small $h$ present, the term $-h\sigma^{x}_{2}$ will break the degeneracy, since 
$\sigma^{x}_{2} \ket{I} = \ket{II}$ and vice-versa. Therefore we end up having a new set of eigenstates $\ket{\pm} = 1/\sqrt{2}(\ket{I} \pm \ket{II})$ with an 
energy splitting of $2h$. Now, since $\bra{-}\sigma^z_{2} \ket{+} = 1$, we see from Eq.~\eqref{Auto_statesz} that for $l = 2$, half the states of the spectrum 
in the autocorrelator will contribute to a oscillatory term $e^{\pm i 2h t}$. 
This exactly explains the oscillations seen in Fig.~\ref{fig:sz_autocorr2} (a). 
Eventually, for later times the oscillations decay as terms of order $h^{2}$ and higher in the energy differences become important.

\begin{figure*}[thb]
\centering
\includegraphics[width=\textwidth]{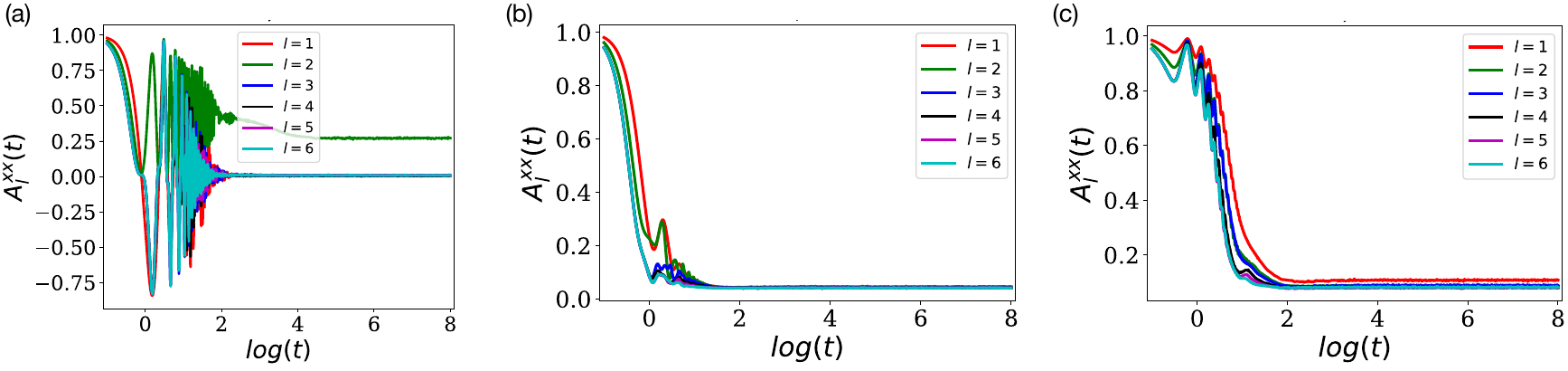}
\caption{Autocorrelation function $A_l^{xx} (t)$ for $L=14$ plotted versus time on a log
scale for different values of the transverse field, (a) $h=0.2$, (b) $h=1$, 
and (c) $h=5$.} \label{fig:sx_autocorr} \end{figure*}

We also note that since $\ket{\pm}$ are eigenstates of $\sigma_2^x$, with
eigenvalues $\pm 1$, these states will contribute to the diagonal terms (i.e., terms with $m=n$ and therefore $E_m = E_n$) in the $xx$-autocorrelator at
$l=2$ in Eq.~\eqref{Auto_statesx}. Since the diagonal terms are time-independent
(as $E_m = E_n$), we expect that the $xx$-autocorrelator at $l=2$
will have a non-zero constant term. This agrees with what we see in
Fig.~\ref{fig:sx_autocorr} (a) for $h=0.2$. 

For large values of $h$, we see in Fig.~\ref{fig:sz_autocorr2} (b) that at several
sites near one end of the system, the $zz$-autocorrelators show pronounced 
oscillations before eventually decaying to zero. All the oscillations have the 
same frequency which is found to be close to $2h$.
We can understand this as follows. For $h \gg 1$, we see from 
Eq.~\eqref{eq:hamil} that the eigenstates of $H$ are given, to lowest order,
by products of eigenstates of $\si_j^x$ for all $j$. An operator $\si_j^z$ 
connects two states which have $\si_j^x = \pm 1$ and therefore unperturbed 
energies equal to $\mp h$. The energy difference of these two states is $2 h$, 
hence Eq.~\eqref{Auto_statesz} implies that the contribution of these two states to the $zz$-autocorrelator at site $j$ will oscillate as 
$e^{\pm i 2h t}$; this explains Fig.~\ref{fig:sz_autocorr2} (b).
Next, we can extend this argument to first order in perturbation theory.
Consider the $zz$-autocorrelator at the first site given by $j=1$ where the
oscillations are most pronounced. To first order 
in the perturbation $V = - \si_1^z \si_2^z \si_2^z$, the two states given by
$|I\ra = |\si_1^x = +1, \si_2^x = a, \si_3^x = b \ra$ and $| II \ra = |\si_1^x = 
-1, \si_2^x = -a, \si_3^x = -b \ra$ will mix (here $a, ~b$ can take values $\pm 
1$). The unperturbed energies of these states are 
$E_I = -h(1+a+b)$ and $E_{II} = h(1+a+b)$ respectively. Hence, to first order in
perturbation theory, the energy of the state lying close to $|I \ra$ will shift 
from $E_I = -h(1+a+b)$ to $E_I' = -h(1+a+b) +1/(E_I - E_{II}) = -h(1+a+b) -1/(2h(1+a+b))$.
Similarly, the perturbation $V$ mixes the two states $|III\ra = |\si_1^x = -1, 
\si_2^x = a, \si_3^x = b \ra$ and $| IV \ra = |\si_1^x = 1, \si_2^x = -a, \si_3^x 
= -b \ra$, and shifts the energy of the state lying close to $| III \ra$ 
from $E_{III} = h(1-a-b)$ to $E_{III}'= h(1-a-b) + 1 /(2h(1-a-b))$. The 
operator $\si_1^z$ connects the states lying close to $|I \ra$ and $| III \ra$, 
and we see from the expressions above that the energy difference between these
two states is 
\bea |E_I' - E_{III}'| &=& 2 h ~+~ \frac{1}{2h} ~\left( \frac{1}{1+a+b} ~+~ 
\frac{1}{1-a-b} \right) \non \\
&=& 2 h ~+~ \frac{1}{h} ~\left( \frac{1}{1-(a+b)^2} \right). \label{eab} \eea

\begin{figure}
\centering
\includegraphics[width=0.43\textwidth]{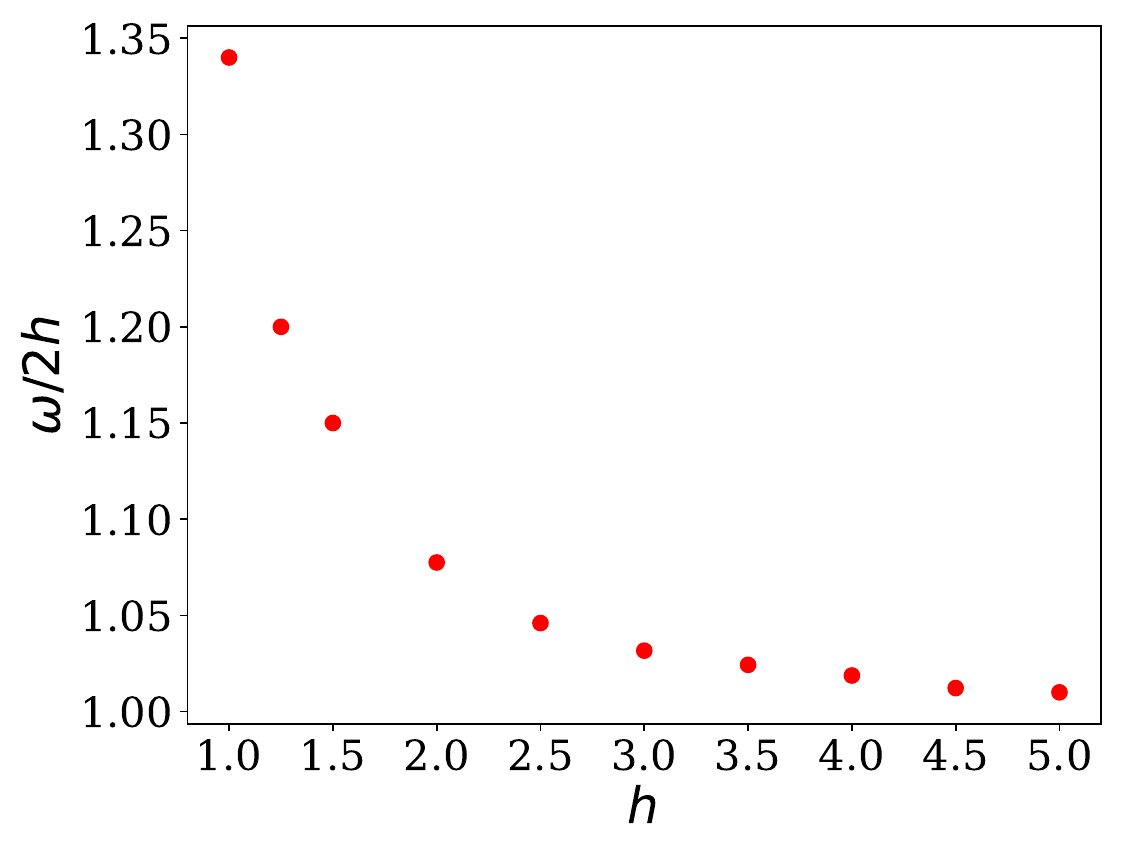}
\caption{Variation of the frequency of oscillations of $A^{zz}_{l=1}(t)$ at the 
end site with the transverse field $h$ for $L=14$. For large $h$ the dependence is consistent with the perturbative result $\omega/(2h) = 1+ 1/(6h^2)$.}
\label{fig:sz_auto_osc_h} \end{figure}

According to Eq.~\eqref{Auto_statesz}, therefore, the oscillations will have the
frequency given in Eq.~\eqref{eab}. Now, since $a, ~b$ can independently take the
values $\pm 1$, giving rise to four possibilities, the expression in 
Eq.~\eqref{eab} can take two possible values given by $2h+(1/h)$ (when $a=-b$)
and $2h-(1/3h)$ (when $a=b$). Hence we expect the oscillations to have a 
frequency $\om$, where $\om/(2h) = 1 + 1/(2h^2)$ and $1 - 1/(6h^2)$. Since these
two cases appear and equal number of times, the average value is given by
$\om/(2h) = 1 + (1/6h^2)$. This is in reasonable agreement with the numerical
result shown in Fig.~\ref{fig:sz_auto_osc_h} for large values of
$h$. We note that since the frequency
$\om$ used in that figure is obtained by calculating the position of the peak
of the Fourier transform of the oscillations in Fig.~\ref{fig:sz_autocorr2} (b), 
the decay of the oscillations leads to a small width around the peak. This width
also turns out to be of the order of $1/h$, and we therefore do not see two 
separate peaks at $\om = 2h+(1/h)$ and $2h-(1/3h)$. Remarkably, these early and intermediate time oscillations in $A_l^{zz} (t)$ persist all the way to $h=1$ (Fig.~\ref{fig:sz_auto_osc_h}) for the boundary site when the critical point is approached from $h>1$, while the
other autocorrelators show a reasonably rapid decay in the neighborhood of the
critical point (see Appendix~\ref{appendixD} for the extraction of the oscillation frequency $\omega$ in Fig.~\ref{fig:sz_auto_osc_h}). 

\section{Discussion}
\label{discussion}

A summary of our main results is as follows. Motivated by the one-dimensional TFIM 
which is one of the best studied integrable models with duality and a quantum 
critical point, we have made a detailed study of a generalization in which
there are Ising interactions between three successive spins (instead of two
successive spins as in the TFIM). We find that the model has a $Z_2 \times Z_2$
symmetry for a system with PBC provided that the system size is a multiple of 3.
This symmetry implies that the system consists of four sectors which are 
decoupled from each other, and this leads to three-fold degeneracies in the 
energy spectrum which involves states from three of the four sectors. Next we 
have discussed the duality of the model between $h$ and $1/h$. While the duality
is straightforward to show for an infinite-sized system, the existence of a 
duality turns out to be a subtle issue for finite-sized systems with PBC. 
We find that exact duality holds only if the system size is {\it not} a multiple
of 3. Next, we make a detailed study of the criticality properties of the model 
at the self-dual point given by $h=1$. Using ED and system sizes up to $L=27$, we
use finite-size scaling to first confirm that there is indeed a critical point at
$h=1$, and then to compute the dynamical critical exponent $z$, the order parameter
exponent $\beta$, the magnetic susceptibility exponent $\ga$, and the correlation
length exponent $\nu$. We find that $z=1$ suggesting that the low-energy sector 
of the model at $h=1$ has conformal invariance. We then determine the central
charge $c$ in two different ways (from the length-dependences of the entanglement 
entropy between two parts of the system and of the ground state energy). We find 
that $c$ is close to 1. We then
observe that although the values of $\beta$, $\ga$ and $\nu$ for the two-spin
and three-spin models are different from each other, the ratios $\beta/\nu$ and
$\ga/\nu$ are the same in the two models. This suggests that there is a weak
universality and the three-spin model lies on the AT line,
just like two copies of the TFIM and the four-state Potts model. All models
on this line are known to have $z=1$, $c=1$, and the same values of $\beta /
\nu = 1/8$ and $\ga /\nu = 7/4$. There is a quantum AT model which has a 
parameter $\lambda$ such that two copies of the TFIM and the four-state 
Potts model correspond to $\lambda = 0$ and 1 respectively. Given our numerically obtained value of $\nu \approx 0.75$ for the three-spin model from ED, we estimate this model
corresponds approximately to $\lambda \approx 0.7$.

To better understand the nature of the criticality at the self-dual point of the three-spin model, we have used the DMRG method for much longer chains, but with open boundaries. The DMRG method using longer chains indeed confirms that $c=1$ at the critical point. However, the analysis of the behavior of the smallest excitation gap at $h_c=1$ and its neighborhood reveals important additive and multiplicative logarithmic corrections. Incorporating these and also comparing to the Ashkin-Teller model at $\lambda=1$, in fact, shows that the critical point in the three-spin model is likely to lie
in the four-state Potts universality class. This claim is further supported by an analysis of the Binder cumulants of the two models. Thus, the three-spin model seems to provide a lattice realization of four-state Potts criticality with a smaller Hilbert space dimensionality of $2^L$ with $L$ sites compared to $4^L$ for the Ashkin-Teller model.   

If the three-spin model and the four-state Potts model indeed lie in 
the same universality class at their critical points, they should have
the same emergent symmetries at that point. We would like to make some
observations in favor of this. First, we have seen that the three-spin model
with PBC has an exact three-fold degeneracy for three-fourth of its
states, namely, states whose momenta differ by multiples of $2\pi/(3d)$
have the same energies (here $d$ is the lattice spacing which we
have set equal to 1 in this paper). At the critical point, the lattice
model is gapless
around $k=0$, $2\pi/(3d)$ and $4 \pi/(3d)$. In the continuum theory which
describes modes with low energies and wave lengths much larger than $d$,
all the gapless modes of the lattice model at the critical point
will get mapped to momenta lying 
around $k=0$. We would therefore expect the conformal field theory to have 
a three-fold degeneracy of the low-lying excitations. We now note that 
this is also 
the case for the four-state Potts model at criticality since that model has
several relevant operators, namely, three with conformal dimensions 
$(1/16,1/16)$, one with $(1/4,1/4)$, and three with $(9/16,9/16)$, and three
marginal operators with conformal dimensions $(1,1)$~\cite{verlinde}. The multiplicities of 3 are expected to lead to three-fold degeneracies in the 
low-lying spectrum~\cite{zou,affleck}.
Second, we have seen that the three-spin model naturally has three 
order parameters, $(m_A,m_B,m_C)$, and in the ordered phase, there are four
possible ground states in which the expectation values of the order 
parameters are proportional to $(1,1,1)$, $(1,-1,-1)$,
$(-1,1,-1)$ and $(-1,-1,1)$; this can be seen most clearly if we consider
the limit $h \to 0$ in Eq.~\eqref{eq:hamil}. These four patterns
of the order parameters form a tetrahedron, and the symmetry group
of a tetrahedron is the permutation group $S_4$ of four objects. Next, we
note that the four-state Potts model also has a $S_4$ symmetry~\cite{verlinde}.
To conclude, the three-spin model at its critical point has the same low-energy 
degeneracies and the same emergent symmetry as the four-state Potts model.

We then studied the energy level spacing statistics in a 
particular symmetry sector of a system with open boundary conditions to determine
if the three-spin model is integrable. We find that the level spacing statistics
has the form of the Gaussian orthogonal ensemble, and hence the model is
non-integrable. Next, we find that the model has an exponentially large number
of mid-spectrum zero-energy states which is consistent with an index theorem; the
number of states grows at least as fast as $2^{L/2}$. Further, we
find that the zero-energy states are of two types which we call Type-I and
Type-II. The Type-I states are special because they are simultaneous
zero-energy eigenstates of the two parts of the Hamiltonian (the three-spin
interaction and the transverse field); hence their wave functions do not change
with $h$ in spite of the energy level spacing in their neighborhood being exponentially small in system size. These states thus violate the ETH and
qualify as quantum many-body scars. We have presented the analytical forms
of some of the Type-I states which show that their number
grows at least linearly with the system size. However we do not know the 
form of the growth more precisely (linear, exponential, or some other
dependence). Finally, we have studied the infinite-temperature
autocorrelation functions
for both $\sigma^x$ and $\sigma^z$ at sites close to one end of a large system
with open boundary conditions. We find that far from the critical point, at 
either $h \ll 1$ or $h \gg 1$, some of the autocorrelators show an anomalous 
behavior in that they show pronounced oscillations and
decay very slowly with time. The time scale of decay is much larger than the
inverse of the energy scales in the Hamiltonian; this is unexpected since the
model is non-integrable. We provide a qualitative understanding of the
oscillations using perturbation theory. However, the reason for a large decay 
time is not yet understood analytically. Furthermore, the autocorrelator
for $\sigma^z$ at the end site
shows persistent oscillations at short and intermediate time scales
even when $h$ is close to the critical point while the other autocorrelators
decay quickly to zero. An analytic understanding of this feature is lacking
as of now. 

So far as experimental realizations of our three-spin model are concerned, it
is known that optical lattices of two atomic species can be a suitable platform to 
generate a variety of spin-1/2 Hamiltonians. In particular, it has been shown
that in a one-dimensional system of triangles formed by an optical lattice, 
the two-spin interactions can be made to vanish by varying the tunneling and  
collisional couplings~\cite{pachos}. This gives rise to an effective Hamiltonian with three-spin interactions of the form $\sigma^{z}_{j} \sigma^{z}_{j+1} 
\sigma^{z}_{j+2}$. Furthermore, a local field magnetic field $\Vec{B}$ can be 
applied which can be tuned by applying appropriately detuned laser fields which 
generate a term of the form $B \sigma_j^x$ in the Hamiltonian. It may also be
feasible to realize effective Hamiltonians with two- and three-spin terms which
are generated by ion-laser interactions in trapped-ion systems~\cite{andrade}. 
A three-spin NMR quantum simulator based on a diethylfluoromalonate molecule has 
been used to experimentally realize a system with two- and three-spin Ising interactions~\cite{peng}.
Turning to studies of the scar states, it may be possible to initialize a system 
in a state which is close to the simplest scar states of the form shown in
Figs.~\ref{fig:typeI} (b) and \ref{fig:typeIc} (b) since these involve 
preparing only neighboring sites to be in spin-singlet or spin-triplet states,
and to then study the time-evolution of such a state.

\vspace{0.6cm}
\centerline{\bf Acknowledgments}
\vspace{0.2cm}

A.S. thanks Hosho Katsura for illuminating discussions. 
D.S. thanks Chethan Krishnan for stimulating discussions in the early stages of 
this work. S.N. thanks Luka Pave\v{s}i\'c and Zala Lenar\v{c}i\v{c} for very 
useful discussions. A.S. and D.S. acknowledge useful discussions with the participants
in the ICTS program titled ``Periodically and quasi-periodically driven complex systems"
(code: ICTS/pdcs2023/6). S.S. and S.N. thank ICTP in Trieste, Italy
for its hospitality during the program titled ``School on Quantum Many-Body Phenomena out of Equilibrium: from Chaos to Criticality (smr 3867)".
S.S. thanks MHRD, India for financial support through 
the PMRF. D.S. acknowledges funding from SERB, India through project JBR/2020/000043. S.N. acknowledges the support by the projects J1-2463, N1-0318 and P1-0044 
program of the Slovenian Research Agency and the QuantERA grant T-NiSQ, by MVZI, QuantERA II JTC 2021. The DMRG calculations were performed at the spinon cluster of 
Jo\v zef Stefan Institute, Ljubljana.

\appendix

\section{Binder cumulant}
\label{appendixB}

Here we present another quantity that shows that the critical behavior of the
three-spin model is different to that of the TFIM.  The Binder cumulant, $U_2$, is defined as~\cite{binder,colin,wang}
\begin{equation} U_2 = C + D \frac{\langle m^4 \rangle}{\langle m^2 \rangle^2}, \label{binder_eq} \end{equation}
where the order parameter $m$ and the normalization constants $C, D$ are defined appropriately for a given model so that $U_2$ has the values 0 and
1 in the thermodynamic limit in the disordered and the ordered phase respectively. For the two-spin TFIM we have $m^2= (\frac{1}{L}\sum_{i}^L\sigma^{z}_i)^2$ with 
$C= 3/2$ 
and $D = -1/2$. For our three-spin model the order parameter is defined as in the Eq.~\eqref{order_parameter} with $C= 5/2$ and $D= 
-3/2$~\cite{sandvik}. For the AT model, $m^2=m_\sigma^2+m_\tau^2$, where
$m_{\sigma}=(1/L)\sum_{i=1}^L \sigma_i^z$ and $m_{\tau}=(1/L)\sum_{i=1}^L \tau_i^z$, for which $C=2$ and $D=-1$~\cite{sandvik}. Furthermore, $\langle O 
\rangle$ in Eq.~\eqref{binder_eq} denotes $\langle \psi_0| O |\psi_0\rangle$, where $|\psi_0\rangle$ equals the ground state at a finite size 
$L$, and $O$ equals either $m^2$ or $m^4$.

\begin{figure}
\centering
\includegraphics[width=0.45\textwidth]{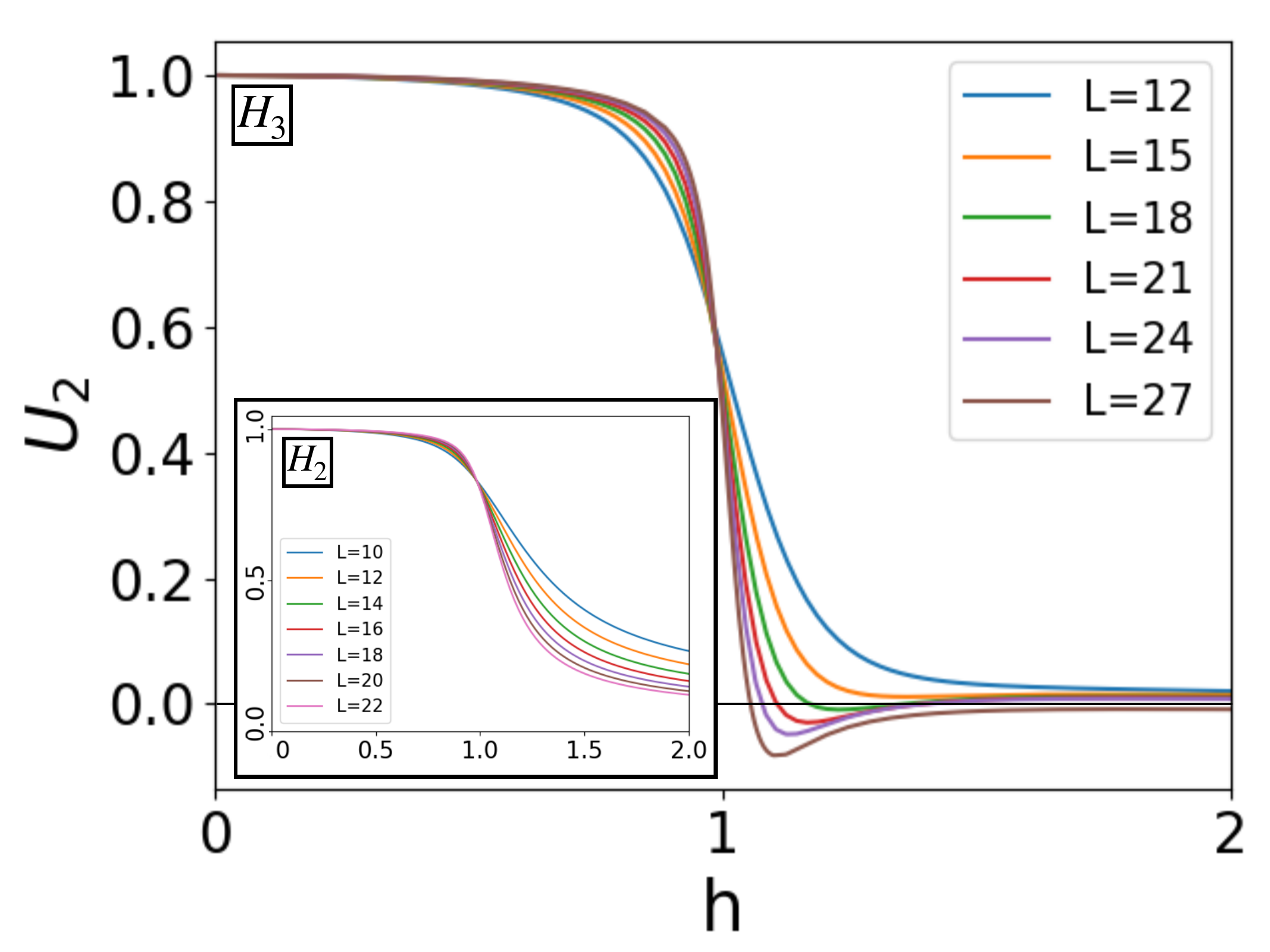}
\caption{Plot of Binder cumulant $U_2$ defined in Eq.~\eqref{binder_eq} as a
function of the field $h$ for the three-spin model. The plots for different 
system sizes cross each other close to $h_c$. We observe a negative dip close 
to $h_c$, the magnitude of which increases with the system size. This is in
contrast to the TFIM where the Binder cumulant is a monotonic function as 
shown in the inset.} \label{fig:binder_long} \end{figure}

We plot the Binder cumulant for the ground state of $H_3$ in Fig.~\ref{fig:binder_long} as a function of the transverse field $h$ using results from ED. As 
expected they cross close to the critical point for different system sizes. More interestingly, for our model, there is a negative dip in $U_2$ close to the 
critical point for $L \geq 18$. The dip increases in magnitude as we go to larger system sizes, however it does not increase faster than $L$; thus the phase transition close to 
$h_c$ is still continuous in nature \cite{jin,capponi, vollmayr}. However 
this is starkly different from the monotonic behavior of $U_2$ for the two-spin 
case as can be seen in the inset of Fig.~\ref{fig:binder_long}.

\begin{widetext}
\begin{figure*}[htb]
\centering
\includegraphics[width=0.98\textwidth]{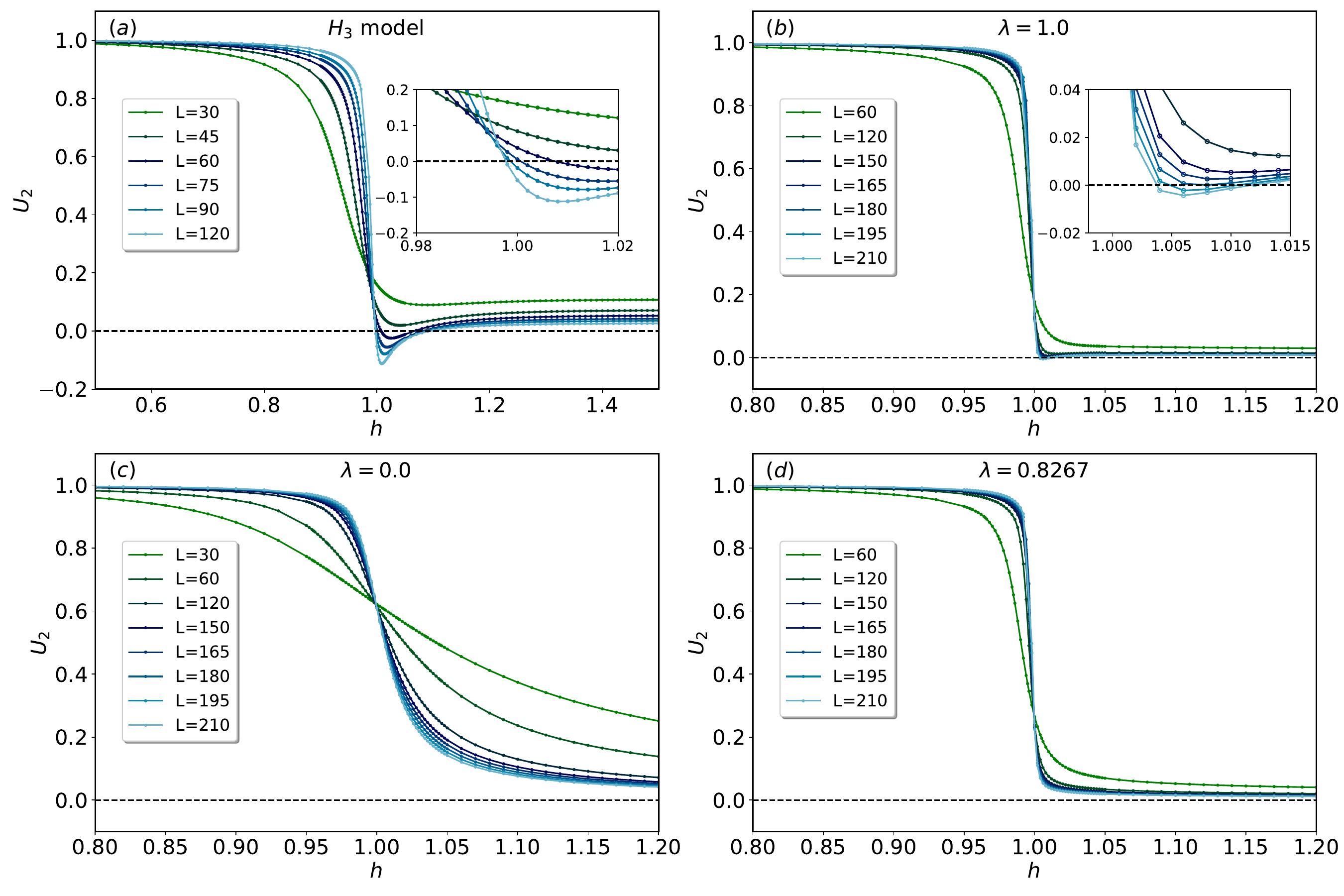}
\caption{Behavior of the Binder cumulant $U_2$ as a function of $h$ for various system sizes $L$ obtained from DMRG on open chains for (a) the 
three-spin model, and for the AT model at couplings (b) $\lambda=1.0$, (c) $\lambda=0$, and (d) $\lambda=0.8267$. The insets of panels (a) and (b) show 
the development of the negative dips in $U_2$ in a clearer manner.}
\label{fig:dmrgBinder} \end{figure*}
\end{widetext}

We now show results for the computation of $U_2$ for longer chains, but with open boundaries, using DMRG both for the three-spin model as well as for the AT 
model at three values of $\lambda$ respectively (see Fig.~\ref{fig:dmrgBinder}). As expected, all the panels show a crossing of 
$U_2$ for different $L$ in the neighborhood of the critical point $h_c=1$. However, only the upper two panels (see Fig.~\ref{fig:dmrgBinder} (a) for the 
three-spin model and Fig.~\ref{fig:dmrgBinder} (b) for the AT model at $\lambda=1$) show a pseudo-first-order behavior, i.e., non-monotonic behavior 
of $U_2$ as a function of $h$ with increasing $L$ and the corresponding negative dip at sufficiently large $L$. The behavior of the negative dips is 
shown more clearly in the insets of both these figures. Fig.~\ref{fig:dmrgBinder} (c) shows the behavior of $U_2$ for the AT model at 
$\lambda=0$ which is equivalent to two decoupled copies of the TFIM. As expected here, $U_2$ does not show any non-monotonic behavior and stays bounded 
between $0$ and $1$. More interestingly, the AT model at the coupling $\lambda=0.8267$, which corresponds to $\nu=0.7233$, does not show any sign of 
non-monotonicity or negative dips as well from the available data with $L \leq 210$.  

\section{Expectation values of local operators and duality for finite systems}
\label{appendixC}

In this appendix, we will study the expectation values of local operators
in all the eigenstates of the Hamiltonian, and we will see that something striking occurs when the system has an exact self-duality. Let us write the
Hamiltonian in Eq.~\eqref{eq:hamil} in the form shown in Eq.~\eqref{hzx}.
It is then interesting to plot the expectation value of, say, $X$ in the 
different eigenstates of $H$ versus the energies of those states. If the system
is integrable, we might expect to see a fragmented kind of pattern corresponding
to the different conserved sectors, while for a non-integrable system, we would
not expect to see any special pattern.

\begin{widetext}
\begin{figure*}[htb]
\centering
\includegraphics[width=0.98\textwidth]{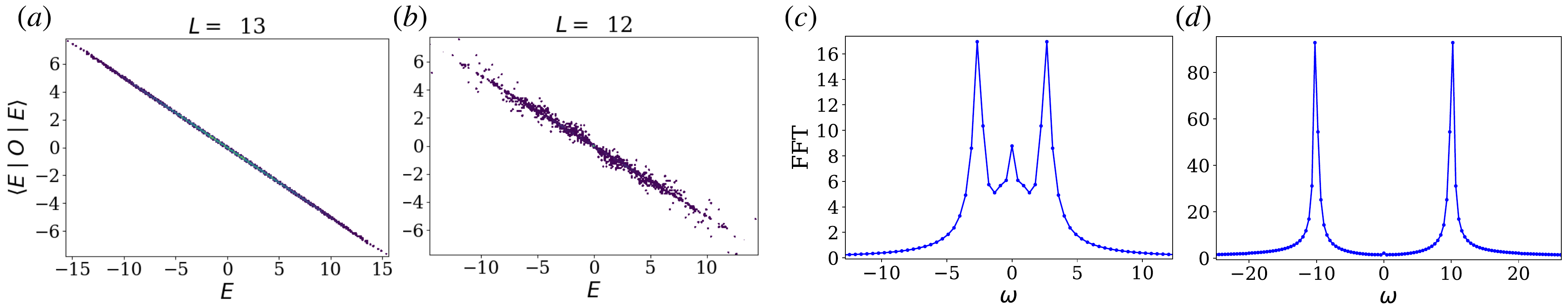}
\caption{(a) Expectation value of the operator $\la \psi_n | 
X | \psi_n \ra$ plotted as a function of the energy $E_n$, where $\psi_n $ is 
the $n$-th energy eigenstate, for $h=1$. We see that for a system size $L=13$
which is not a multiple of three, the curve falls exactly a straight line with 
a slope of $-0.5$. (b) The same plot for a system size $L=12$ which is a
multiple of three. We see that there is a scattering of the date points as 
the system no longer has exact self-duality. (c) $\tilde{A}^{zz}_{l=1} (\omega)$ at $h=1$ for a system with $L=14$. (d) The same quantity at $h=5$.} \label{fig:op_spread_FFT_autocor} \end{figure*}
\end{widetext}

It turns out that something interesting happens at the critical point $h=1$.
We have discussed in Sec.~\ref{duality} that the system with $L$ sites and 
PBC is self-dual only if $L$ is not a multiple of 3.
We therefore expect, from Eq.~\eqref{virial2}, that a plot of $\la \psi_n | 
X | \psi_n \ra$ versus $E_n$ (where $n$ denotes the eigenstate number) should
be a perfect straight line with slope equal to $-1/2$ if $L$ is not a multiple 
of 3. But if $L$ is a multiple of 3, the self-duality does not hold and the 
plot is not expected to be a perfect straight line; we expect several points 
to lie away from the straight line. This is exactly what we see in
Fig.~\ref{fig:op_spread_FFT_autocor} (a) and (b). We see a perfect straight line for a system
size $L=13$ where there is exact self-duality but a scattering of points
when $L=12$ where the self-duality is not exact.


\section{Fourier transform of the autocorrelator $A^{zz}$ for $h \ge 1$}
\label{appendixD}

Here we present the plots for $\tilde{A}^{zz}_{l=1} (\omega)$ for $h =1 $ and $5$, which are obtained by taking the Fourier transform of $A^{zz}_{l=1}(t)$ at those 
values of $h$ over a large interval $\tau$. More precisely,
\begin{equation} \tilde{A}^{zz}_{l=1} (\omega) ~=~ 
\frac{1}{\tau} ~\int_{\tau_0}^{\tau_0+\tau} dt ~e^{-i\omega t} A^{zz}_{l=1}(t),
\end{equation}
where the starting time
$\tau_0$ is chosen carefully to eliminate the initial decay so 
that it can capture the oscillatory nature of the correlator. The time interval $\tau$ is taken to be large enough to contain a certain number of complete 
oscillations which provides a better resolution in the frequency space. Since the oscillations persisted more for larger $h$, the values of $\tau$ also was taken to 
be different for different values of $h$. In our case $\tau$ was chosen to be $15$ for $h=1$ and $20$ for $h=5$. The well-defined peaks denote the frequencies of 
oscillations of the autocorrelator shown in Fig.~\ref{fig:sz_auto_osc_h}. Moreover, the decays of the oscillations lead to a finite width around the peak of 
the Fourier transform which goes roughly as $1/h$. This is clear from the figure, as the width is much smaller for $h = 5$ (Fig.~\ref{fig:op_spread_FFT_autocor} (d)) 
than for $h=1$ (Fig,~\ref{fig:op_spread_FFT_autocor} (c)). The small peak around $\omega =
0$ for $h=1$ is because of the initial high value of the autocorrelator which adds 
to a constant value while performing the Fourier transform over a time interval.

\end{document}